\documentclass{aastex631}

\usepackage{bbold}
\usepackage{booktabs}
\usepackage{multirow}

\shorttitle{StarCLR: Contrastive Learning Representation for Astronomical Light Curves}
\shortauthors{Ding et al.}
\graphicspath{{./}}

\begin{document}

\title{StarCLR: Contrastive Learning Representation for Astronomical Light Curves}

\author{Junyao Ding}
\affiliation{Department of Physics, College of Science, Tibet University, Lhasa 850000, China}
\affiliation{CAS Key Laboratory of Optical Astronomy, National Astronomical Observatories, Chinese Academy of Sciences, Beijing 100101, China}
\affiliation{Research Center for Scientific Data Hub, Zhejiang Laboratory, Hangzhou 311100, China}

\author{Xiaodian Chen}
\email{chenxiaodian@nao.cas.cn}
\affiliation{CAS Key Laboratory of Optical Astronomy, National Astronomical Observatories, Chinese Academy of Sciences, Beijing 100101, China}
\affiliation{School of Astronomy and Space Science, University of the Chinese Academy of Sciences, Beijing, 100049, China}
\affiliation{Institute for Frontiers in Astronomy and Astrophysics, Beijing Normal University,  Beijing 102206, China}

\author{Xinyi Gao}
\affiliation{CAS Key Laboratory of Optical Astronomy, National Astronomical Observatories, Chinese Academy of Sciences, Beijing 100101, China}
\affiliation{School of Astronomy and Space Science, University of the Chinese Academy of Sciences, Beijing, 100049, China}

\author{Xiaoyu Tang}
\affiliation{Research Center for Scientific Data Hub, Zhejiang Laboratory, Hangzhou 311100, China}

\author{Shu Wang}
\affiliation{CAS Key Laboratory of Optical Astronomy, National Astronomical Observatories, Chinese Academy of Sciences, Beijing 100101, China}

\author{Yang Huang}
\affiliation{School of Astronomy and Space Science, University of the Chinese Academy of Sciences, Beijing, 100049, China}

\author{Xinyu Qi}
\affiliation{Research Center for Scientific Data Hub, Zhejiang Laboratory, Hangzhou 311100, China}

\author{Guirong Xue}
\affiliation{Research Center for Scientific Data Hub, Zhejiang Laboratory, Hangzhou 311100, China}

\author{Ali Luo}
\affiliation{CAS Key Laboratory of Optical Astronomy, National Astronomical Observatories, Chinese Academy of Sciences, Beijing 100101, China}
\affiliation{School of Astronomy and Space Science, University of the Chinese Academy of Sciences, Beijing, 100049, China}

\author{Jifeng Liu}
\affiliation{CAS Key Laboratory of Optical Astronomy, National Astronomical Observatories, Chinese Academy of Sciences, Beijing 100101, China}
\affiliation{School of Astronomy and Space Science, University of the Chinese Academy of Sciences, Beijing, 100049, China}



\begin{abstract}

With the rapid development of time-domain surveys, the availability of massive light curve data offers new opportunities for studying stellar evolution and variable star classification, while simultaneously posing challenges for feature extraction and modeling. We present StarCLR, a contrastive pretraining framework for large-scale light curves. By constructing positive pairs from partially overlapping sub-sequences, StarCLR encourages the model to learn temporal representations. We pretrain StarCLR on the TESS dataset and fine-tune it for variable star classification on three surveys with distinct observational characteristics, namely TESS (18 types), ZTF (11 types), and Gaia (24 types). StarCLR achieves macro-$F_1$ scores of 84.35\%, 87.82\%, and 92.73\%, and micro-$F_1$ scores of 94.46\%, 92.83\%, and 99.49\%, respectively. Compared with LSTM and Transformer trained from scratch, StarCLR performs better on TESS and ZTF, with the largest gains on sparsely sampled ZTF light curves, demonstrating promising generalization. For Gaia, which involves a broader class space, the evaluation is not directly comparable, and performance is likely influenced by astrophysical features, resulting in a more limited contribution from the pretrained backbone. Systematic ablations on embedding design, pooling strategy, and pretraining settings further indicate that the pretrained representations provide performance gains by capturing informative temporal characteristics of light curves. Looking ahead, with standardized datasets and more diverse labeling schemes, the generalization ability of StarCLR can be further enhanced. \footnote{Inference code and pretrained model weights available at \url{https://github.com/dj-y/StarCLR-Inference}}

\end{abstract}



\section{Introduction} \label{sec:intro}

The systematic study of variable stars dates back to the late nineteenth and early twentieth centuries. Since the widespread adoption of CCD technology in the 1990s, rapidly growing survey data have strongly accelerated this field. The Optical Gravitational Lensing Experiment \citep[OGLE,][]{OGLE}, operating for more than three decades, has identified nearly 900,000 variable stars and substantially advanced variable-star evolution and classification studies \citep{OGLE_cite1, OGLE_cite2}. Kepler provided high-precision photometry and nearly 200,000 high-quality light curves \citep{Kepler}; based on about 150,000 targets, \citet{Kepler_cite1} performed a systematic periodic-variability analysis. The All-Sky Automated Survey for Supernovae \citep[ASAS-SN,][]{ASAS_SN} has released about 680,000 variable-star light curves since 2013 \citep{ASAS_SN_Number}, and \citet{ASAS_AN_cite1} further built a catalog of about 379,000 stars in eight major classes from roughly 55 million isolated sources. The Transiting Exoplanet Survey Satellite \citep[TESS,][]{TESS} further expanded this effort \citep{TESS_Tic}: during its two-year primary mission, \citet{TESS_cite1} identified nearly 46,000 high-confidence periodic variables from two-minute cadence data, plus about 38,000 moderate-confidence candidates, while \citet{GaoClassification} classified about 72,000 periodic variables into 12 classes. Up to Data Release 23, the Zwicky Transient Facility \citep[ZTF,][]{ZTF1,ZTF2} produced roughly 5 billion light curves, among which \citet{ChenClassification} discovered and classified around 780,000 periodic variables. Gaia Data Release 3 \citep[Gaia DR3,][]{GaiaDR3} published light curves for about 11.75 million variable sources, classified into 25 principal categories through its dedicated variability pipeline \citep{GaiaDR3_classification}.

As observational datasets continue to grow, manual inspection and heuristic rules are no longer sufficient. Machine learning has therefore become central to variable-star classification, enabling efficient processing and finer categorization of massive light-curve datasets. Early machine-learning pipelines mainly depended on handcrafted features derived from prior knowledge, such as periods from power spectra or Fourier-decomposition parameters. These features improved interpretability and helped control overfitting. A representative approach is the Gaussian Mixture Model \citep[GMM,][]{GMM}, which models data as a weighted sum of Gaussian components and infers labels from posterior probabilities in feature space. Using the HCTSA package \citep{GMM_cite1_feature} to extract 6,492 time-series features, \citet{GMM_cite1} trained a GMM on 1,319 stars and reported about 82\% balanced accuracy across seven classes. Another widely used method is Random Forest \citep{RF}, an ensemble of decision trees with majority voting. Its threshold-based splits align with astronomy practice, where classes are often separated by period, amplitude, and related parameters (e.g., Cepheids versus RR Lyrae; Delta Scuti versus High-Amplitude Delta Scuti). Following this idea, \citet{RF_cite1} trained two Random Forest models for variable/non-variable discrimination and RR Lyrae identification, ultimately producing a catalog of 71,755 RR Lyrae stars from 155 million sources.

Although traditional machine-learning methods have achieved notable progress, they still depend heavily on handcrafted features. Designing and validating such features requires substantial domain expertise and tuning. As data volume grows, this dependence limits scalability and adaptability. Deep learning offers an end-to-end alternative by learning features directly from raw light curves, typically with better scalability and generalization. For example, \citet{CNN_cite1} proposed a hybrid CNN--LSTM framework that extracts features from both light curves and frequency spectra, reaching 99.1\% recall in binary EB versus non-EB classification on 58,768 training sources. \citet{CNN_cite2} introduced a multi-resolution design using global and informative subcurve/sub-spectrum inputs: dominant and local spectral peaks were used to fold light curves, and CNNs then performed feature extraction and classification. On a combined ZTF+TESS+Kepler dataset with 19,641 training samples, this method achieved 87\% accuracy across 17 classes.

These results show that deep learning improves light-curve classification, but supervised training still requires large labeled datasets. In the broader machine-learning community, self-supervised pretraining has substantially reduced label dependence. For example, SimCLR \citep{simCLR} achieved 85.8\% top-5 accuracy on ImageNet when fine-tuned with only 1\% labeled data. Motivated by these advances, several pretraining–fine-tuning frameworks have recently emerged in astronomy to better exploit large-scale unlabeled light curve data and to improve data efficiency in downstream tasks. \citet{Astromer, Astromer2} developed Astromer, trained analogously to Bidirectional Encoder Representations from Transformers \citep[BERT;][]{BERT}: random observations are masked and reconstructed for self-supervised learning. After pretraining, Astromer supports few-shot fine-tuning and remains competitive with only 20, 100, or 500 labeled sources. Complementing masked reconstruction, \citet{FALCO} proposed FALCO with a Generative Pre-trained Transformer~2-like \citep[GPT-2;][]{GPT2} autoregressive objective that predicts the next time step from preceding observations. Pretrained on Kepler light curves, FALCO achieved 95\% accuracy for eight-class variable-star classification, 0.1305~dex RMSE for surface-gravity estimation, and 87\% precision for flare identification. Beyond unimodal pretraining, \citet{astrom3} introduced $\mathrm{AstroM}^3$, a contrastive multimodal self-supervised framework that jointly embeds photometric time series, spectra, and astrophysical features. It achieved 91.5\% accuracy on time-series photometry classification and also supports misclassification identification, similarity search, and anomaly detection.

Despite these advances, most astronomical pretraining--fine-tuning methods are still survey-specific and often require separate models for different datasets. To address this limitation, we propose StarCLR, a contrastive learning framework designed to learn temporal representations that yield measurable classification performance when applied to datasets that differ from the pretraining data.

The paper is organized as follows. Section~\ref{sec:data} introduces the datasets used for pretraining and fine-tuning, along with the corresponding preprocessing procedures. Section~\ref{sec:method} describes the proposed methodology, including the problem definition, the design of input and position embeddings, the attention mechanism module, the pretraining objective, and the overall model architecture. Section~\ref{sec:result} presents the experimental results, including the pretraining loss curves and the quantitative evaluation of downstream classification performance. Section~\ref{sec:discussion} provides a detailed discussion, where we compare different model architectures and training strategies, analyze the effects of pretraining and feature ablation, examine the impact of different pretraining datasets on downstream performance, and investigate the role of embedding design, pooling strategies, and representation structure through UMAP analysis. Finally, Section~\ref{sec:conclusion} summarizes this work and outlines prospects for future research.

\section{Data} \label{sec:data}

This section describes the datasets used for pretraining and fine-tuning in StarCLR, together with their processing procedures. Pretraining learns general temporal representations from large-scale unlabeled light curves, so we use publicly released TESS light curves. For fine-tuning, existing catalog classifications are used as labels and matched to corresponding light curves. We collect variable-star samples from TESS, ZTF, and Gaia DR3 to evaluate model performance across surveys with different sampling conditions.

\subsection{Pretrain data} \label{subsec:PretrainData}

The pretraining light curves were drawn from the first 69 TESS sectors. To help the model learn representative temporal features, we applied selection criteria to retain likely variable-star candidates and reduce contamination from non-variable or noise-dominated light curves. The filtering procedure was as follows:

\begin{enumerate}
\item \textbf{Flux filtering}: all PDC-corrected flux values (PDCSAP\_FLUX) within the light curve were required to be positive.
\item \textbf{Outlier removal}: all measurements deviating by more than \(5\sigma\) were discarded to reduce the influence of extreme outliers on training.
\item \textbf{Periodicity analysis}: a Lomb–Scargle periodogram \citep{Lomb,Scargle} was computed, with the false alarm probability (FAP) required to be less than 0.01 to ensure statistical significance of periodic signals.
\item \textbf{Fourier decomposition}: each light curve was fitted with a fourth-order Fourier model, and the coefficient of determination \(R^2\) was required to exceed 0.1.
\end{enumerate}

To enforce a uniform sequence length in PyTorch training \citep{pytorch}, all light curves were processed into tensors of length 8192. This length covers most TESS time steps while remaining compatible with GPU memory constraints. The TESS data used in this work have cadences of 2, 10, and 30 minutes, with each sector spanning about 27.4 days; consequently, sequence lengths range from about one thousand to over twenty thousand time steps. Figure~\ref{fig:LightCurve}(a) shows one representative TESS light curve. Sequences shorter than 8192 were zero-padded, while longer ones were randomly cropped into two subsequences for data augmentation, improving robustness to different observational segments. In total, 764,986 light curves were retained for pretraining.

\subsection{Fine-tuning data} \label{subsec:FineTuningData}

The class labels used for fine-tuning are obtained from existing survey catalogs of TESS, ZTF, and Gaia. These catalogs are typically generated through automated or semi-automated classification pipelines rather than human-verified ground truth on a per-source basis, and may therefore contain certain levels of noise and systematic biases.

\subsubsection{TESS data} \label{subsubsec:FineTuningTESSData}

For the TESS fine-tuning stage, we adopted the original variable star labels provided by \citet{GaoClassification}, which were produced through a supervised random forest–based classification pipeline. Each target was traced back to its corresponding light curve using the TESS Input Catalog (TIC) identifier, and the same filtering criteria as in the pretraining stage were strictly applied to ensure consistency and reliability of the training samples. The fine-tuning dataset covers 18 distinct types of variable stars, and the TESS panel of Table~\ref{tab:AllTuningData} summarizes the detailed statistics for the TESS dataset, including variable types, brief descriptions, and the corresponding number of targets.

For data processing, all light curves were again standardized to tensors of length 8192. Unlike pretraining, no random cropping was applied during fine-tuning. Instead, sequences longer than 8192 time steps were truncated to their first 8192 points to avoid potential leakage from data augmentation. This prevents different slices of the same light curve from appearing in both training and validation/test sets, preserving evaluation independence. In total, 19,796 light curves were collected for fine-tuning.

\subsubsection{ZTF data} \label{subsubsec:FineTuningZTFData}

For the ZTF fine-tuning stage, we used the variable star catalog provided by \citet{ChenClassification}, generated through an auto-manual classification pipeline based on Density-Based Spatial Clustering of Applications with Noise \citep[DBSCAN;][]{DBSCAN}. Unlike TESS, ZTF is a ground-based survey and therefore exhibits distinct characteristics in terms of temporal sampling, photometric bands, and measurement precision. Accordingly, we designed a data processing workflow tailored to the properties of ZTF when constructing the fine-tuning dataset.

Specifically, we randomly selected time-series observations from half of the fields in ZTF DR23 and cross-matched them with the variable-star catalog to obtain labeled $g$-band light curves. To ensure reliable periodic variability, the following criteria were applied: 

\begin{enumerate}
\item \textbf{Periodicity analysis}: a Lomb–Scargle periodogram was computed, requiring \(\text{FAP} < 0.01\)
\item \textbf{Fourier decomposition}: a fourth-order Fourier model was fitted to each light curve, with the goodness of fit required to satisfy \(R^2 > 0.4\)
\item \textbf{Outlier removal}: only measurements with \(\text{catflags} \ne 32768\) were retained, and all points deviating by more than \(3\sigma\) were discarded.
\end{enumerate}

The resulting ZTF fine-tuning dataset covers 11 classes of variable stars and comprises a total of 242,169 targets, as summarized in the ZTF panel of Table~\ref{tab:AllTuningData}. The ZTF light curves contain highly variable numbers of observations, typically ranging from a few hundred to several thousand time steps. As illustrated in Figure~\ref{fig:LightCurve}(b), the sampling cadence is coarser than that of TESS but denser than Gaia. To standardize the input, all sequences were truncated or padded to a fixed length of 1024, a choice that balances capturing most variability features with computational efficiency.

\subsubsection{Gaia data} \label{subsubsec:FineTuningGaiaData}

The Gaia fine-tuning dataset was derived from the variable star catalog released in Gaia DR3, which includes 24 classes of variable sources encompassing both periodic and partially non-periodic variability. The catalog classifications were produced by a dedicated pipeline composed of multiple machine-learning algorithms. We excluded all measurements with \(\text{variability\_flag\_g\_reject} = \text{true}\) and applied a \(3\sigma\) outlier removal to the light curves. To further ensure the reliability of labels, we selected only candidates from the \texttt{gaiadr3.vari\_classifier\_result} table with classification probabilities greater than 75\%; for certain classes—namely AGN, LPV, DSCT\textbar GDOR\textbar SXPHE, ECL, RR, RS, S, and SOLAR\_LIKE—we adopted a more stringent threshold of 95\%. The final training set contained 377,333 targets, as summarized in the Gaia panel of Table~\ref{tab:AllTuningData}.

\begin{figure}[ht!]
\centering
\includegraphics[width=0.7\textwidth]{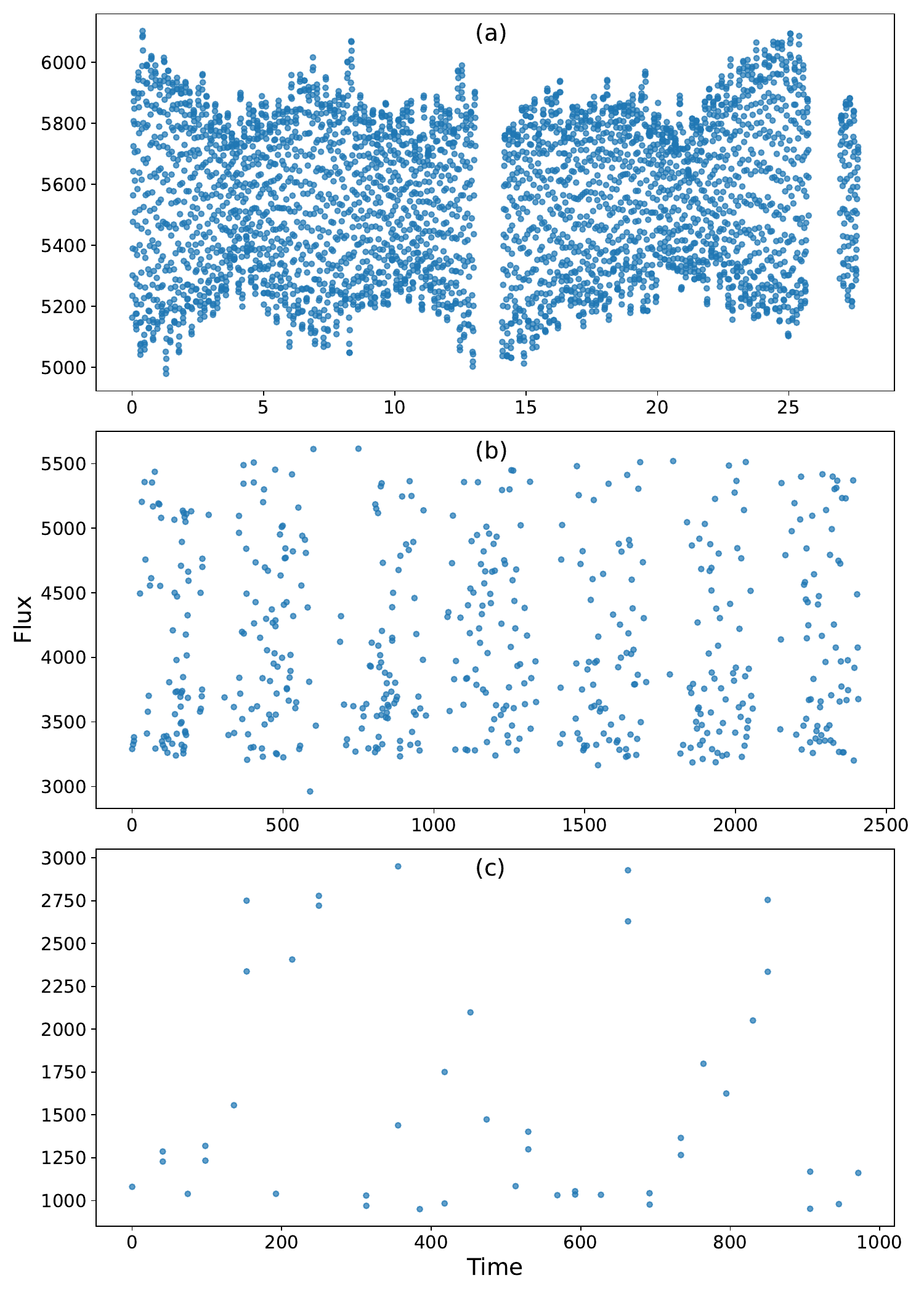}
\caption{Examples of light curves from three surveys: (a) TESS, (b) ZTF, and (c) Gaia. The horizontal axis is the observation time with the minimum value subtracted, and the vertical axis is the flux. The TESS light curve spans a single sector ($\sim$27.4 days), the shortest baseline among the three, whereas ZTF and Gaia extend to several thousand days. In terms of cadence, TESS has the shortest time interval between successive observations, followed by ZTF. However, Gaia provides the sparsest sampling overall, with long gaps due to its scanning law.}
\label{fig:LightCurve}
\end{figure}

Unlike TESS and ZTF, no periodicity-based pre-filtering is applied to Gaia. As a result, the dataset covers a broader and partially out-of-distribution class space, and the evaluation is not directly comparable to TESS and ZTF. Because Gaia provides relatively few observations per source, with the number of time steps typically ranging from a few dozen to a few hundred, all light curves were standardized to tensors of maximum length 256. As illustrated in Figure~\ref{fig:LightCurve}(c), the sampling is sparse due to the scanning law. Sequences shorter than 256 were zero-padded, while longer sequences were truncated to the first 256 time steps.

\begin{longtable}{llr}
\caption{Summary of variable star types and total object counts in the TESS, ZTF, and Gaia datasets.}
\label{tab:AllTuningData}\\

\toprule
\textbf{Type} & \textbf{Description} & \textbf{Total Objects} \\
\midrule
\endfirsthead

\toprule
\textbf{Type} & \textbf{Description} & \textbf{Total Objects} \\
\midrule
\endhead

\midrule
\multicolumn{3}{r}{\emph{Continued on next page}} \\
\midrule
\endfoot

\bottomrule
\endlastfoot

\multicolumn{3}{c}{\textbf{TESS}} \\
\midrule
DSCT   & $\delta$ Scuti-type pulsators & 6,069 \\
EA     & Algol-type eclipsing binaries & 6,024 \\
EW     & W Ursae Majoris-type contact binaries & 3,186 \\
EB     & $\beta$ Lyrae-type semi-detached binaries & 933 \\
ROT    & Rotational variables & 922 \\
UV     & UV Ceti-type flare stars & 640 \\
RRab   & RR Lyrae variables (fundamental mode) & 411 \\
DCEP   & Classical Cepheids (fundamental mode) & 278 \\
HADS   & High-amplitude $\delta$ Scuti variables & 246 \\
RRcd   & RR Lyrae variables (first overtone and double-mode) & 211 \\
SPB    & Slowly pulsating B-type stars & 179 \\
YSO    & Young stellar objects & 176 \\
GDOR   & $\gamma$ Doradus-type pulsators & 147 \\
BCEP   & $\beta$ Cephei-type pulsators & 96 \\
UG     & U Geminorum-type dwarf novae & 95 \\
DCEPS  & Cepheids (first overtone mode) & 81 \\
GCAS   & $\gamma$ Cassiopeiae-type eruptive variables & 73 \\
T2CEP  & Type II Cepheids & 29 \\
\midrule
Total  &  & 19,796 \\
\midrule\midrule

\multicolumn{3}{c}{\textbf{ZTF}} \\
\midrule
EW     & W Ursae Majoris-type contact binaries & 158,141 \\
EA     & Algol-type eclipsing binaries & 22,192 \\
RSCVN  & RS Canum Venaticorum-type binaries & 19,077 \\
RR     & RR Lyrae variables (fundamental mode) & 11,683 \\
SR     & Semiregular pulsating variables & 9,391 \\
BYDra  & BY Draconis-type rotational variables & 6,864 \\
DSCT   & $\delta$ Scuti-type pulsators & 6,267 \\
RRc    & RR Lyrae variables (first overtone mode) & 4,682 \\
Mira   & Mira Ceti-type long-period variables & 3,232 \\
CEP    & Classical Cepheids (fundamental mode) & 513 \\
CEPII  & Type II Cepheids & 127 \\
\midrule
Total  &  & 242,169 \\
\midrule\midrule

\multicolumn{3}{c}{\textbf{Gaia}} \\
\midrule
LPV          & Long-period variables & 106,834 \\
ECL          & Eclipsing binaries & 80,886 \\
AGN          & Active galactic nuclei & 70,876 \\
S            & Short-timescale variable stars & 40,549 \\
YSO          & Young stellar objects & 15,104 \\
RR           & RR Lyrae variables & 11,469 \\
ELL          & Ellipsoidal variables & 11,361 \\
DSCT$\vert$GDOR$\vert$SXPHE & $\delta$ Scuti, $\gamma$ Doradus, and SX Phoenicis-type pulsators & 10,020 \\
RS           & RS Canum Venaticorum-type binaries & 9,497 \\
SOLAR\_LIKE  & Solar-like variable stars & 5,347 \\
BE$\vert$GCAS$\vert$SDOR$\vert$WR & B-type emission-line stars, $\gamma$ Cas, S Doradus, Wolf--Rayet & 3,986 \\
CEP          & Cepheid variables & 3,440 \\
ACV\textbar CP\textbar MCP\textbar ROAM\textbar ROAP\textbar SXARI & Alpha2 Canum Venaticorum, Chemical Peculiar,    &    \\
             & Rapidly Oscillating Am/Ap stars, SX Arietis variables & 3,008    \\
CV           & Cataclysmic variables & 2,515 \\
SN           & Supernovae & 734 \\
SDB          & Subdwarf B-type pulsators & 468 \\
WD           & White dwarf variables & 265 \\
SPB          & Slowly pulsating B-type stars & 239 \\
BCEP         & $\beta$ Cephei-type pulsators & 238 \\
EP           & Exoplanet transiting systems & 214 \\
SYST         & Symbiotic variables & 152 \\
ACYG         & $\alpha$ Cygni-type pulsators & 69 \\
MICROLENSING & Microlensing events & 35 \\
RCB          & R Coronae Borealis-type variables & 27 \\
\midrule
Total        &  & 377,333 \\
\end{longtable}

\subsubsection{Supplementary features} \label{subsubsec:SupplementaryFeatures}

During the pretraining stage, light curves are processed with separate treatments for the time and flux dimensions, allowing the model to focus primarily on relative temporal structures and variability patterns rather than absolute scales. This design encourages the model to learn characteristic features of variable stars from their light curve morphology, while information related to absolute physical quantities is not explicitly emphasized at this stage. To complement this limitation, we introduce additional astrophysical features during fine-tuning to enhance the model’s sensitivity to stellar physical properties. These features are mainly drawn from Gaia DR3 and include:

\begin{enumerate}
\item \textbf{Period}: Variability period is a key physical indicator for distinguishing different types of variable stars. We computed the period using the Lomb–Scargle method.
\item \textbf{Amp}: The peak-to-valley amplitude, defined as \(\frac{\max(\text{flux})}{\min(\text{flux})}\), recovers the intensity of brightness variations that is diminished during flux normalization.
\item \textbf{Parallax, Parallax\_error}: Parallax and its uncertainty provide information on source distances and measurement reliability, both of which are crucial for estimating intrinsic luminosities.
\item \textbf{BP–RP}: Color index that serves as a key parameter for Hertzsprung–Russell (HR) diagram analysis.
\item \textbf{Gmag}: The mean magnitude in the G band, offering brightness information.
\item \textbf{Wesenheit G magnitude}: To mitigate the effect of interstellar extinction on brightness measurements, we computed the Wesenheit magnitude defined as \(G - 1.89 \times (B_P - R_P) - 5 \log_{10}(100 \times D)\), where \(D\) is the distance estimated from parallax (in kpc) and the extinction coefficient is from \citet{2019ApJ...877..116W}. 
\end{enumerate}

The set of supplementary features differed across datasets: for TESS we used \{Period, Amp, Wesenheit G, BP–RP, Parallax\}; for ZTF and Gaia we used \{Period, Amp, Parallax, Parallax\_error, BP–RP, Gmag\}. Note that some Gaia sources lack period or parallax measurements. To explicitly signal these missing cases to the model, we applied special negative placeholders: a value of –1 was used when the period was unavailable, and –10 was used for missing parallax or parallax error. As these values lie far outside any physically reasonable range, the model can readily interpret them as “missing indicators” rather than true astrophysical quantities during training.

\subsection{Preprocessing} \label{subsec:Preprocessing}

During preprocessing, we apply different normalization strategies to the time and flux dimensions. For the time axis, we remove the absolute time offset by shifting each light curve to start at zero,
\begin{equation}
t' = t - \min(t)
\end{equation}
so that the model becomes invariant to the starting epoch of observations. For the flux dimension, we apply min--max normalization to map values into $[0,1]$,
\begin{equation}
f' = \frac{f - \min(f)}{\max(f) - \min(f)}
\end{equation}
where $t'$ and $f'$ denote the preprocessed time and flux values, respectively.

This design removes the arbitrary absolute starting time of each light curve while preserving the overall observing baseline and the relative time intervals between measurements. In contrast, min–max normalization of flux mitigates survey-specific photometric scales and instrumental zero-point offsets, encouraging the model to focus on relative variability patterns rather than absolute flux values.

Compared with using absolute timestamps directly, the time-shifted representation reduces the influence of absolute time offsets on model training and typically improves training stability. Although sampling cadences may vary across different datasets, this preprocessing strategy provides a more consistent temporal reference frame for representation learning. Any light-curve information that may be attenuated by these normalizations can be partially reintroduced during fine-tuning by concatenating supplementary astrophysical features with the learned embeddings, which can further improve downstream classification performance.

\section{Method} \label{sec:method}

This section introduces StarCLR for learning representations from irregularly sampled astronomical light curves. StarCLR combines the global sequence-modeling capability of Transformers with contrastive objectives to capture temporal patterns across surveys with different cadences. We first define the input representation and problem setting, then describe the embedding and attention designs for irregular time series, followed by the hierarchical contrastive pretraining objective and the overall network architecture.

\subsection{Problem definition} \label{subsec:definition}

StarCLR is designed to process single-band light curve data. Let a set of light curves be denoted as \(L = \{l_1, l_2, \cdots, l_N\}\), containing \(N\) samples in total. Each light curve \(l_i\) has a dimension of \(T_i \times 2\), where \(T_i\) is the number of time steps for the \(i\)-th curve, and the two features correspond to the observation time and the associated flux.  

The model learns a parameterized nonlinear function \(f_\theta\) that maps each light curve \(l_i\) to a high-dimensional representation \(r_i\), capturing temporal and variability patterns in the sequence. Specifically, \(r_i = \{r_{i,1}, r_{i,2}, \cdots, r_{i,T_i}\}\), where each \(r_{i,t} \in \mathbb{R}^H\) is the hidden state at time step \(t\), and \(H\) is the hidden dimension. This stepwise extraction captures both local and global variability patterns, providing representations for downstream tasks such as classification.  

\subsection{Position Embedding} \label{subsec:PE}

StarCLR adopts the positional encoding formulation in \citet{PE} and uses a sinusoidal scheme similar to Transformer \citep{transformer}, with a key modification for irregularly sampled time series. In the original Transformer, positional embeddings are built from discrete token indices, which implicitly assume uniform sampling. However, sinusoidal positional encoding is intrinsically continuous in position, and its discreteness comes only from using integer indices.

Astronomical light curves are indexed with continuous timestamps and are typically sampled irregularly. To better capture this characteristic, we replace the discrete position indices with the corresponding observation times and scale them by a factor of $2\pi$, so that the positional embeddings vary continuously with time. A comparison of different positional encoding designs is presented in Section~\ref{subsec:embeddingAnalysis}. This modification preserves the functional form of sinusoidal positional encoding while naturally extending its applicability to irregularly sampled continuous time domains.

Formally, we define a set of angular frequencies as
\begin{equation}
\omega_i = \frac{2\pi}{10000^{2i/H}},
\end{equation}
and construct the positional encoding for an observation time $t$ as
\begin{equation}
p_{t,2i} = \sin(\omega_i t), \quad
p_{t,2i+1} = \cos(\omega_i t),
\end{equation}
where $H$ denotes the dimensionality of the positional embedding.

\subsection{Attention block} \label{subsec:attention}

StarCLR employs multi-head attention to enhance extraction of temporal features from light curves. This mechanism computes multiple self-attention heads in parallel, enabling the model to capture diverse patterns from different subspaces and better represent temporal dependencies. Compared with CNN or RNN, attention directly models dependencies between distant time steps, making it well suited for long-range correlations across multiple variability cycles.  

Self-attention dynamically computes a weighted representation for each position based on its correlation with all other time steps, thereby modeling global dependencies. Multi-head attention extends this by projecting input features into multiple subspaces, where self-attention is computed independently for finer-grained extraction and parallel modeling. Each attention head \(head_i\) is defined as follows:  

\begin{equation} \label{eq:head}
    \text{head}_i = \text{softmax}\left( \frac{Q_i K_i^\top}{\sqrt{d_k}} \right) V_i
\end{equation}

where \(Q_i\), \(K_i\), and \(V_i\) are the query, key, and value matrices obtained by applying linear transformations to the input \(X\):  

\begin{equation}
    Q_i=XW_i^Q, \; K_i=XW_i^K, \; V_i=XW_i^V
\end{equation}

with \(W_i^Q\), \(W_i^K\), and \(W_i^V\) denoting the learnable parameter matrices for the \(i\)-th attention head. The dimension \(d_k\) corresponds to the size of the key and query vectors, typically defined as \(d_k = H / k\), where \(H\) is the hidden dimension of the input and \(k\) is the number of attention heads. The scaling factor \(\sqrt{d_k}\) alleviates issues of vanishing or exploding gradients, thereby improving training stability.  

The outputs of the \(k\) attention heads are concatenated and linearly transformed to produce the final attention output:  

\begin{equation}
    \text{MultiHead}(Q,K,V)=\text{Concatenate}(\text{head}_1,\text{head}_2,\cdots, \text{head}_k)W^O
\end{equation}

where \(W^O\) is a learnable projection matrix for integrating the head outputs. This design enables the model to simultaneously capture information from multiple feature subspaces, thereby strengthening its capability for sequence modeling and representation learning.  

\subsection{Pretrain loss} \label{subsec:loss}

The contrastive learning strategy adopted in this work follows \citet{TS2Vec}, whose central idea is to construct positive and negative pairs such that the model pulls positive pairs closer in the representation space while pushing negative pairs apart, thereby enhancing the discriminability and generalization of the learned representations.  

Prior to being fed into the model, each light curve undergoes data augmentation to generate positive pairs. Specifically, the original light curve is duplicated, and two augmented versions are obtained through random cropping, with the constraint that the resulting subsequences maintain partial overlap. This ensures that the two views remain physically consistent but differ in form, thus constituting a positive pair.

To capture multi-level associations across sequences, we employ a \textit{hierarchical contrastive loss} consisting of two complementary components: the \textit{instance-wise contrastive loss}, which enforces separability across different sources, and the \textit{temporal contrastive loss}, which enhances the model’s sensitivity to intra-source temporal dynamics.  

\subsubsection{Instance-wise contrastive loss} \label{subsubsec:instance_loss}

The core idea of the instance-wise contrastive loss is to guide the model to learn a representation space in which different augmented views of the same instance (positive pairs) are close to each other, while representations from different instances (negative pairs) are pushed apart. Let the batch size be \(B\). In one forward and backward pass, we have a batch of input light curves \(L = \{l_1, l_2, \cdots, l_B\}\). After augmentation, the set expands to \(L = \{l_1, l'_1, l_2, l'_2, \cdots, l_B, l'_B\}\). Feature extraction yields the representation set \(R = \{r_1, r'_1, r_2, r'_2, \cdots, r_B, r'_B\}\), where \(r_i\) is defined in Section~\ref{subsec:definition}. The instance-wise contrastive loss for the \(i\)-th light curve at time step \(t\) is formulated as:  

\begin{equation}
    \ell^{(i,t)}_{inst} = -\log \frac{\exp(r_{i,t} \cdot r'_{i,t})}{\sum_{j=1}^{B} (\exp(r_{i,t} \cdot r'_{j,t}) + \mathbb{1}_{[ i \neq j ]} \exp(r_{i,t} \cdot r_{j,t}))}
\end{equation}

where \(\mathbb{1}\) denotes the indicator function used to exclude self-comparisons, defined as: 

\begin{equation}
    \mathbb{1}_A(x) =
    \left\{
    \begin{array}{ll}
        1, & \text{if } x \in A \\ 
        0, & \text{else}
    \end{array}
    \right.
\end{equation}

\subsubsection{Temporal contrastive loss} \label{subsubsec:temporal_loss}

The temporal contrastive loss aims to capture feature differences along the temporal dimension. For the two augmented representations \(r_i\) and \(r'_i\) of the same light curve, this loss compares feature vectors across different time steps within the same instance, thereby strengthening the model’s sensitivity to temporal evolution patterns. Formally, the temporal contrastive loss for the \(i\)-th light curve at time step \(t\) is defined as:  

\begin{equation}
    \ell^{(i,t)}_{temp} = -\log \frac{\exp(r_{i,t} \cdot r'_{i,t})}{\sum_{t' \in \Omega} (\exp(r_{i,t} \cdot r'_{i,t'}) + \mathbb{1}_{[t \neq t']} \exp(r_{i,t} \cdot r_{i,t'}))}
\end{equation}

where \(\Omega\) denotes the set of overlapping time indices between the two augmented sequences. 

\subsubsection{Hierarchical contrastive loss} \label{subsubsec:hierarchicalLoss}

The two loss functions compute similarity along complementary dimensions: the instance-wise loss focuses on inter-sequence discriminability, while the temporal loss emphasizes intra-sequence temporal dynamics. Together, they form the dual contrastive loss, defined as:  

\begin{equation}
    \mathcal{L}_{\text{dual}} = \frac{1}{B \cdot |\Omega|} \sum_{i=1}^{B} \sum_{t \in \Omega} \left( \ell^{(i,t)}_{temp} + \ell^{(i,t)}_{inst} \right)
\end{equation}

where \(|\Omega|\) is the length of the overlapping index set, i.e., the number of time steps involved in contrastive learning.  

\begin{figure}[ht!]
\centering
\includegraphics[width=0.65\textwidth]{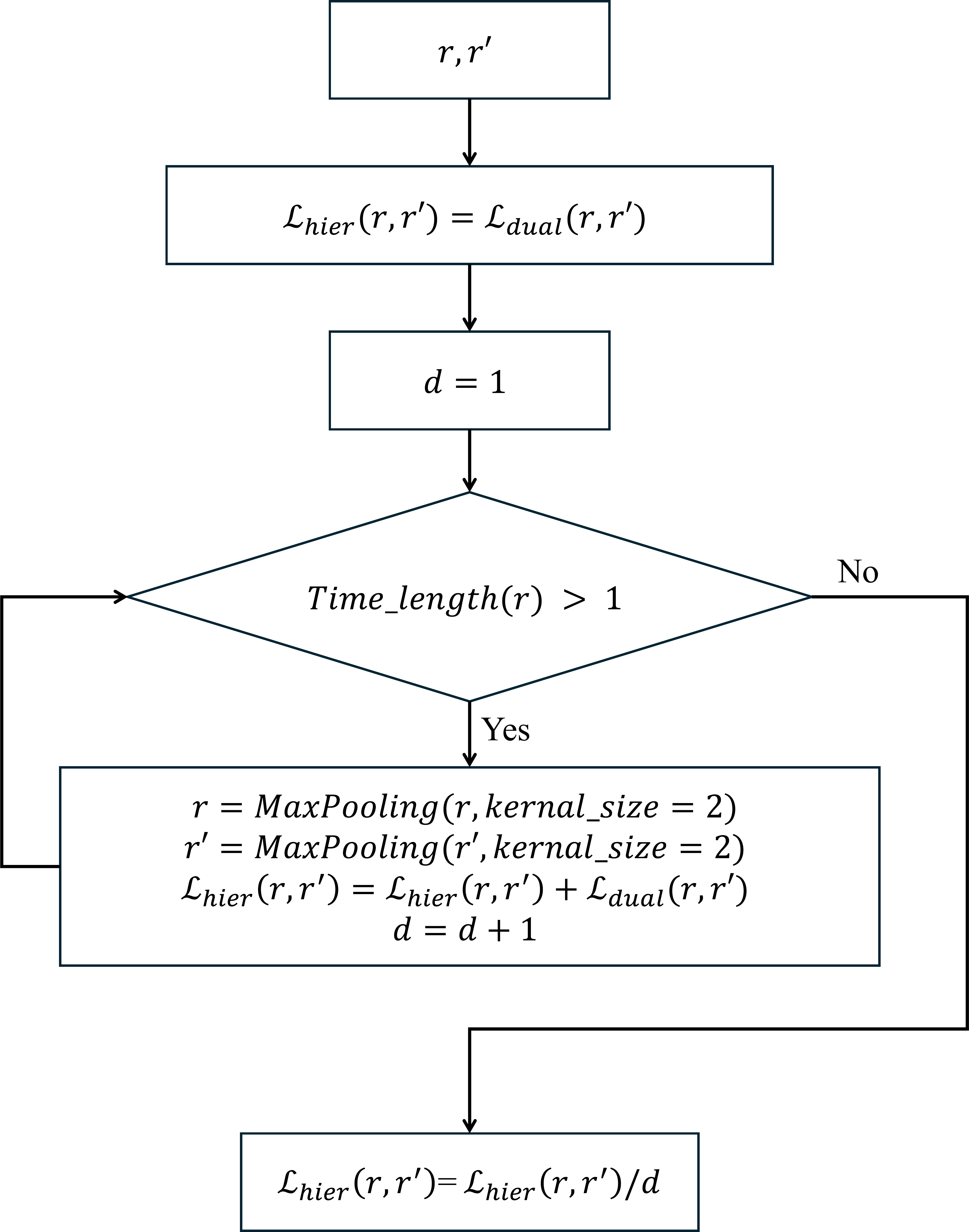}
\caption{Workflow of hierarchical contrastive loss computation.}
\label{fig:HierarchicalLoss}
\end{figure}

To further improve robustness and multi-scale awareness, we adopt a hierarchical contrastive loss. At each layer, a max-pooling operation with kernel size 2 is applied to compress sequence length and capture contextual information at different scales. This allows the model to jointly learn similarities at both the global sequence level and the local temporal level, thereby enhancing the robustness and multi-scale sensitivity of the learned representations. The overall process is illustrated in Figure~\ref{fig:HierarchicalLoss}.  

\subsection{Model architecture} \label{subsec:architecture}

The backbone of StarCLR adopts an encoder-only Transformer architecture \citep{BERT,transformer}. Specifically, our implementation is adapted from the BERT model provided by the HuggingFace Transformers library \citep{transformers}, with modifications to accommodate irregularly sampled astronomical time-series data.

The input to the encoder consists of two-dimensional time-series data composed of observation time and flux. Although the Transformer architecture natively supports variable-length input sequences, the PyTorch implementation requires that all samples within a batch share the same sequence length. During pretraining, the model supports a maximum input length of 8192 time steps: for light curves longer than this limit, 8192 data points are randomly sampled, while shorter sequences are zero-padded to the fixed length. For each raw light curve of length $T_i$, we first align the length. The flux values are then projected to a 512-dimensional feature space through a feed-forward layer. In parallel, the corresponding observation times are mapped to 512-dimensional positional embeddings using the sinusoidal positional encoding described in Section~\ref{subsec:PE}. The two embeddings are combined by element-wise addition, resulting in an input embedding matrix of size $8192 \times 512$.

\begin{figure}[ht!]
\centering
\includegraphics[width=0.6\textwidth]{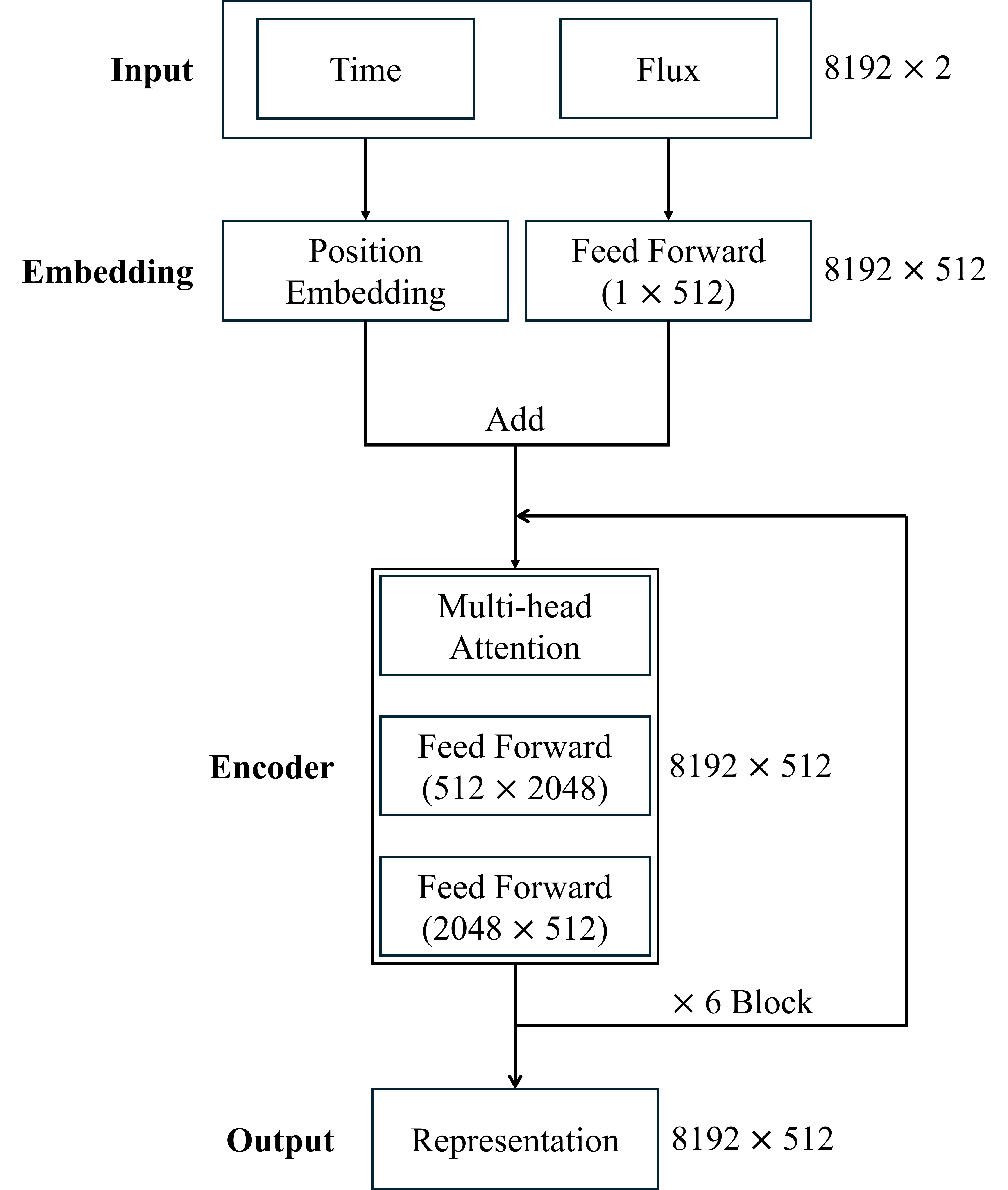}
\caption{The architecture of StarCLR. The input consists of preprocessed time and flux values. The flux sequence is projected into a 512-dimensional feature space through a feed-forward layer, while the corresponding time are encoded into 512-dimensional positional embeddings using the sinusoidal positional encoding. The two representations are combined via element-wise addition to form an input embedding matrix of size \(8192 \times 512\). The resulting embeddings are then passed into a Transformer encoder consisting of six stacked blocks. Each block contains a multi-head attention mechanism and a position-wise feed-forward network, together with residual connections and layer normalization following the standard Transformer design. The final output is a light curve representation of dimension \(8192 \times 512\). The length 8192 corresponds to the pretraining setting, while the Transformer architecture naturally supports variable-length inputs.}
\label{fig:ModelArchitecture}
\end{figure}

To enhance the extraction of temporal features, StarCLR stacks six Transformer attention blocks. Each block consists of a multi-head attention module with eight heads, followed by a position-wise feed-forward network that expands the hidden dimension to 2048 before projecting it back to 512. Residual connections and layer normalization are applied as in the standard Transformer design. The overall model architecture is illustrated in Figure~\ref{fig:ModelArchitecture}.  

\begin{figure}[ht!]
\centering
\includegraphics[width=0.5\textwidth]{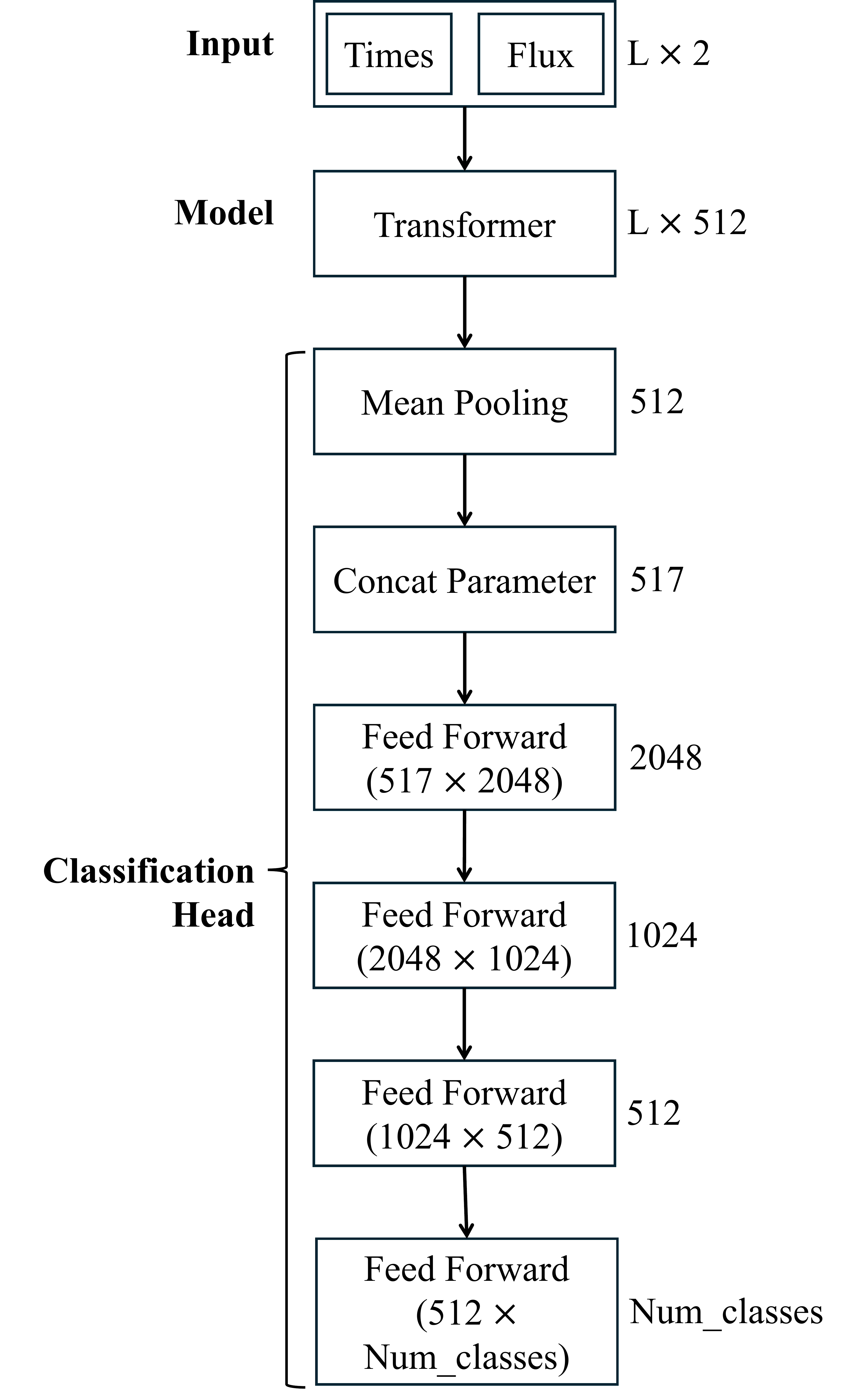}
\caption{StarCLR fine-tuning architecture. The input light curve is represented as an \(L \times 2\) matrix, where \(L\) is the number of time steps and each step consists of time and flux. The sequence length is standardized according to the survey characteristics; for example, 8192 for TESS, 1024 for ZTF, and 256 for Gaia. The backbone is a Transformer encoder that produces an output of dimension \(L \times 512\). Subsequently, mean pooling is applied to obtain a 512-dimensional global representation, which is then concatenated with the supplementary physical features to form a \((512 + d_{\text{sup}})\)-dimensional vector. Finally, this vector is passed through a four-layer feed-forward neural network, yielding the predicted probabilities over variable-star classes.}
\label{fig:ModelFineTuningArchitecture}
\end{figure}

During pretraining, the model relies solely on the contrastive learning objective to acquire feature representations. In the fine-tuning stage, a classification head is added on top of the Transformer backbone to adapt the model to the supervised task of variable star classification. Specifically, the time-step representations produced by the Transformer encoder are aggregated via mean pooling, where all hidden states are averaged to form a global sequence representation. This strategy is robust for handling light curves of variable length and irregular sampling, while also mitigating the impact of noise and outliers. A detailed analysis of different pooling strategies is provided in Section~\ref{subsec:poolingAnalysis}. The resulting global representation is then concatenated with supplementary stellar attributes to construct a joint feature vector, thereby integrating light-curve features with static physical properties. Finally, this vector is passed through a four-layer feed-forward neural network with ReLU activations between hidden layers, yielding the predicted probabilities for each variable star class. The fine-tuning architecture is shown in Figure~\ref{fig:ModelFineTuningArchitecture}.

\section{Result} \label{sec:result}

Based on the datasets in Section~\ref{sec:data} and the model framework in Section~\ref{subsec:architecture}, this section presents StarCLR results for both pretraining and fine-tuning. We focus on contrastive pretraining with TESS light curves. We then apply the pretrained model to TESS, ZTF, and Gaia, fine-tuning only the classification head while keeping the backbone frozen. This setting allows us to evaluate the model’s performance on downstream tasks constructed from different survey datasets. Additional experiments with alternative pretraining sources are discussed in Section~\ref{subsubsec:otherPretrain}.

\subsection{Pretrain} \label{subsec:pretrain}

As described in Section~\ref{subsec:loss}, StarCLR is pretrained with hierarchical contrastive learning to learn temporal light-curve representations. The contrastive loss is applied at the representation layer shown in Figure~\ref{fig:ModelArchitecture}. Pretraining converges stably and reaches a final test loss of 0.02136, providing an effective initialization for fine-tuning. A comparison with non-hierarchical contrastive pretraining is presented in Section~\ref{subsubsec:noHierarchicalPretrain}.

To satisfy the requirement that positive pairs share overlapping temporal segments, we apply structured input augmentation. For each original light curve, after standard preprocessing, we generate two partially overlapping subcurves via a “duplication + random cropping” strategy. If four time points \(t_1 < t_2 < t_3 < t_4\) are randomly selected from the normalized sequence, the original curve \(l\) is cropped into:

\begin{itemize}
    \item \(l_1\): covering the interval \([t_1, t_3]\)
    \item \(l_2\): covering the interval \([t_2, t_4]\)
\end{itemize}

\begin{figure}[ht!]
\centering
\includegraphics[width=1\textwidth]{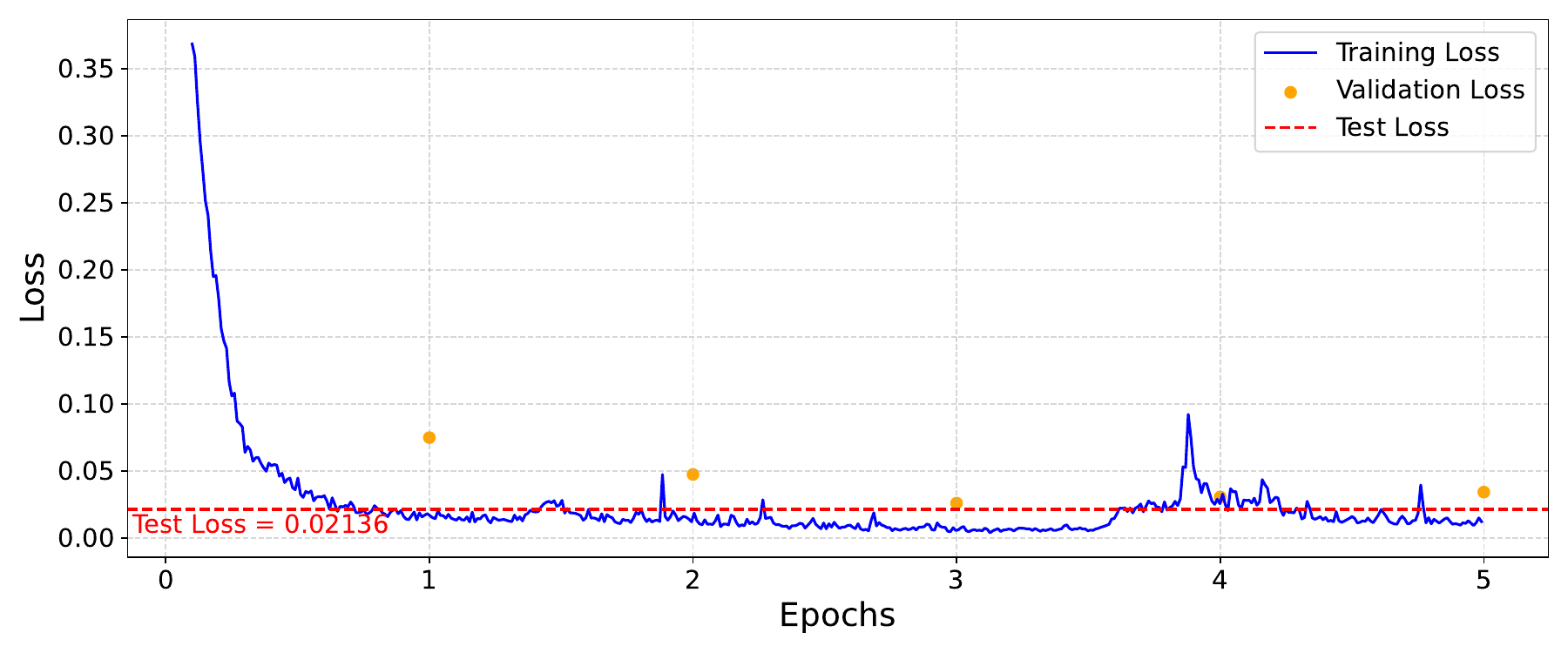}
\caption{Loss curves during the pretraining stage on the TESS dataset. The horizontal axis denotes the training epochs, while the vertical axis represents the loss value. Both training and validation losses are shown as a function of epochs, and the model achieves its optimal performance at around the third epoch.}
\label{fig:PretrainLoss}
\end{figure}

Thus, the intersection \([t_2, t_3]\) is the overlap shared by \(l_1\) and \(l_2\). After passing through the Transformer encoder, representations of overlapping observations from the two subcurves are encouraged to be as similar as possible, forming positive pairs for contrastive learning. Using this strategy, we conduct large-scale pretraining on TESS, enabling the model to learn informative representations of relative temporal variations. The evolution of pretraining loss is shown in Figure~\ref{fig:PretrainLoss}. As introduced in Section~\ref{subsec:PretrainData}, 764,986 TESS light curves were split into training, validation, and test sets at a ratio of 7:1:2. Model selection was based on validation loss, and the checkpoint with the lowest validation loss—at the third epoch—was chosen as the final pretrained model. This model attains a test loss of 0.02136.

\subsection{Fine-tuning} \label{subsec:fineTuning}

The primary objective of fine-tuning is to evaluate whether the pretrained Transformer backbone learns effective temporal representations of light curves. To this end, we keep the backbone frozen and fine-tune only the classification head, so downstream evaluation mainly reflects representation quality rather than survey-specific adaptation, while also reducing computational cost compared with full fine-tuning. StarCLR achieves competitive performance across surveys, with macro-$F_{1}$ scores of $84.35\% \pm 1.32\%$, $87.82\% \pm 0.29\%$, and $92.73\% \pm 0.46\%$ on TESS, ZTF, and Gaia, respectively, and corresponding micro-$F_{1}$ scores of $96.46\% \pm 0.27\%$, $92.83\% \pm 0.06\%$, and $99.49\% \pm 0.01\%$. The detailed evaluation protocol and impact of different classification heads are discussed in Section~\ref{subsubsec:pretrainAblation}. Unless otherwise specified, all reported classification results are averaged over at least three independent runs with different random seeds, and uncertainties correspond to the standard deviation across runs.

To evaluate performance across classes, we also compute Precision, Recall, and $F_1$-score from the confusion matrix:

\begin{equation}
    \text{Precision}=\frac{\text{True Positive}}{\text{True Positive} + \text{False Positive}}
\end{equation}

\begin{equation}
    \text{Recall}=\frac{\text{True Positive}}{\text{True Positive} + \text{False Negative}}
\end{equation}

\begin{equation}
    F_1 = 2\times\frac{\text{Precision} \times \text{Recall}}{\text{Precision} + \text{Recall}}
\end{equation}

For multi-class classification, we report both macro- and micro-$F_1$ scores. Macro-$F_1$ is computed by first calculating the $F_1$-score for each class and then averaging equally across classes, treating all classes with the same importance. In contrast, micro-$F_1$ is computed by aggregating total true positives, false positives, and false negatives over all classes before computing $F_1$, and is therefore dominated by frequent classes.

\begin{figure}[ht]
\centering
\includegraphics[width=\textwidth]{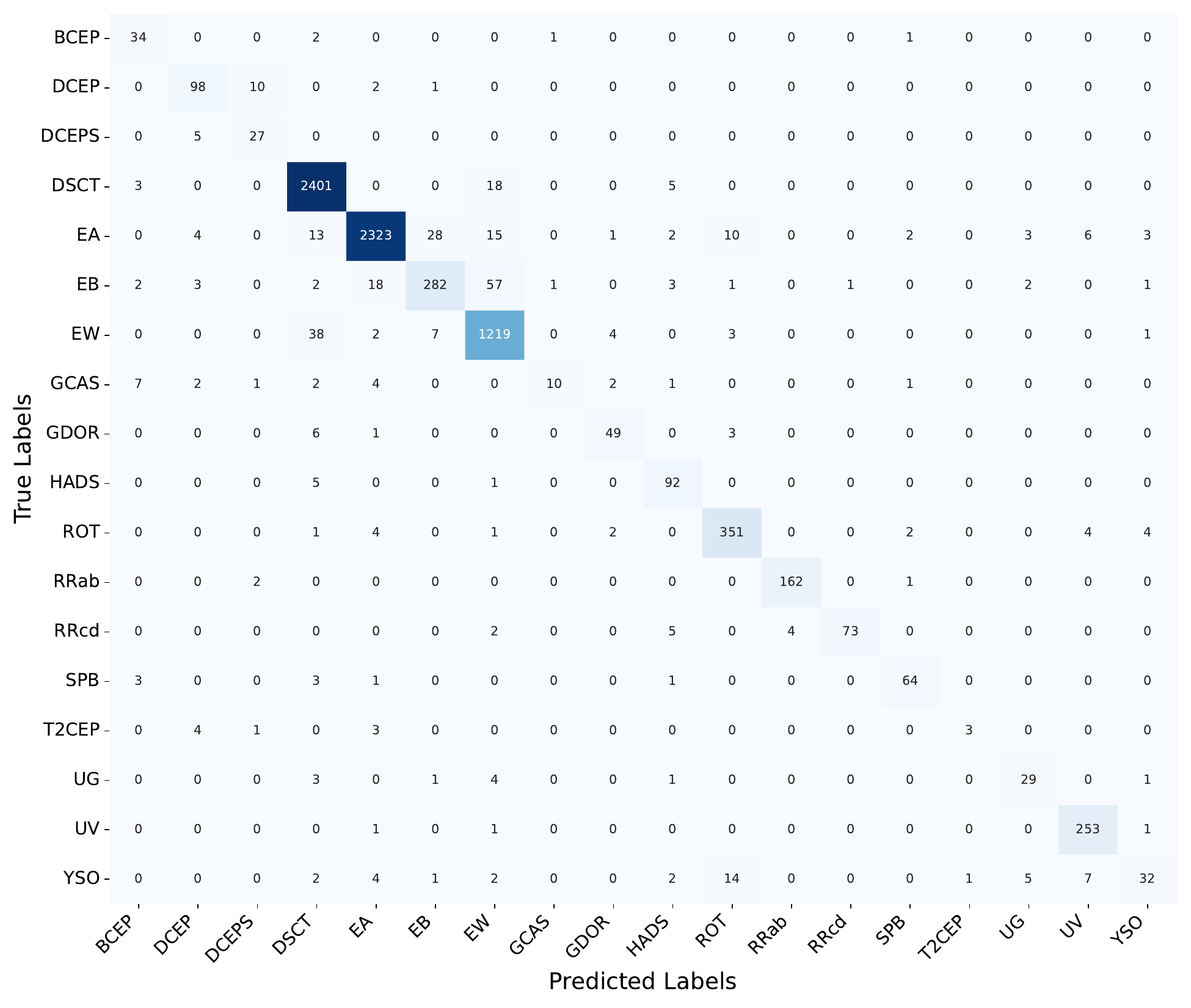}
\caption{Confusion matrix of the StarCLR model on the TESS fine-tuning task. Rows correspond to the true labels and columns to the predicted labels.}
\label{fig:TESSConfusionMatrix}
\end{figure}

As described in Section~\ref{subsec:FineTuningData}, the TESS dataset was split into 9,898 samples for training, 1,980 for validation, and 7,918 for testing. The top panel of Table~\ref{tab:PerClassAll} summarizes the classification performance on the TESS test set. StarCLR achieves a macro $F_1$ score of $84.35\% \pm 1.32\%$ and a micro-$F_1$ score of $96.46\% \pm 0.27\%$. The corresponding confusion matrix is shown in Figure~\ref{fig:TESSConfusionMatrix}. It can be seen that the model performs consistently well for classes with sufficient samples, such as DSCT, EA, and EW, achieving $F_1$ scores above $95\%$. However, noticeable confusion remains among classes with highly similar features, including the eclipsing-binary subclasses EA, EB, and EW; the Cepheid subclasses DCEP, DCEPS, and T2CEP; and the low-amplitude pulsators DSCT, BCEP, and SPB. On the other hand, the performance for minority classes is considerably lower. For example, the $F_1$ scores for GCAS reach $50.57\% \pm 4.15\%$, for YSO reach $66.61\% \pm 8.64\%$, and for T2CEP reach $46.56\% \pm 6.33\%$. These results indicate that the model’s ability to identify rare variable types with complex variability patterns remains limited.

\begin{figure}[ht]
\centering
\includegraphics[width=\textwidth]{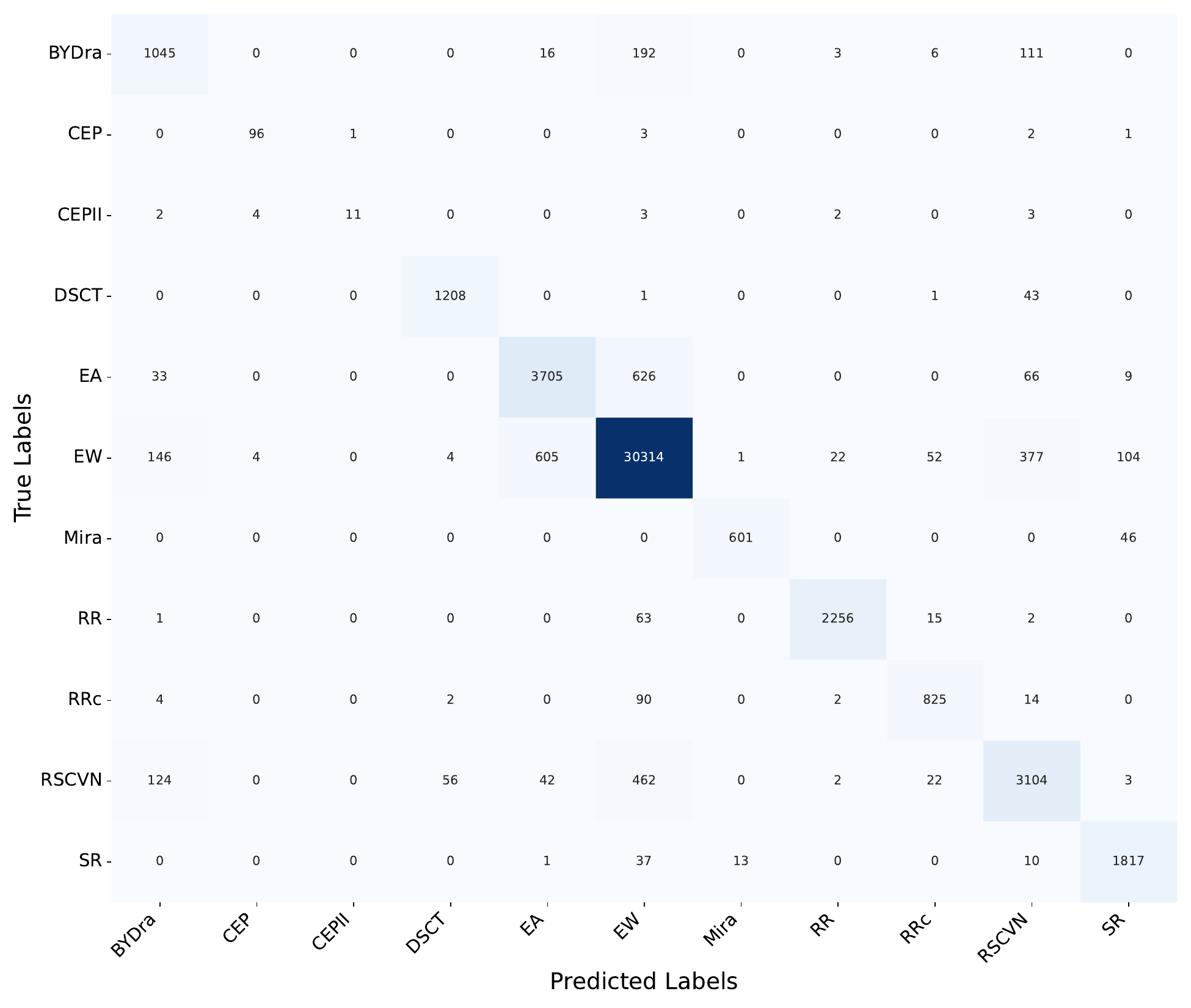}
\caption{Confusion matrix of the StarCLR model on the ZTF fine-tuning task.}
\label{fig:ZTFConfusionMatrix}
\end{figure}

The ZTF dataset was divided into 169,518 training instances, 24,215 for validation, and 48,436 for testing. The middle panel of Table~\ref{tab:PerClassAll} summarizes the classification performance on the ZTF test set. StarCLR achieves a macro $F_1$-score of $87.82\% \pm 0.29\%$ and a micro-$F_1$ score of $92.83\% \pm 0.06\%$. The corresponding confusion matrix is shown in Figure~\ref{fig:ZTFConfusionMatrix}. For well-represented classes, such as EW, EA, DSCT, and RR, the model again shows stable and strong performance. Misclassifications are primarily concentrated in closely related subclasses or underrepresented categories, such as the confusion between EA and EW, the difficulty in separating RR and RRc, and the confusion between CEP and CEPII. In particular, CEPII, which is severely underrepresented, shows the weakest performance with an $F_1$-score of $63.06\% \pm 3.12\%$. Overall, while rare and closely related subclasses remain difficult, performance on ZTF is robust. Importantly, these results demonstrate that a backbone pretrained exclusively on TESS data can generalize effectively to ground-based surveys with distinct observing conditions and cadences.

\begin{figure}[ht]
\centering
\includegraphics[width=\textwidth]{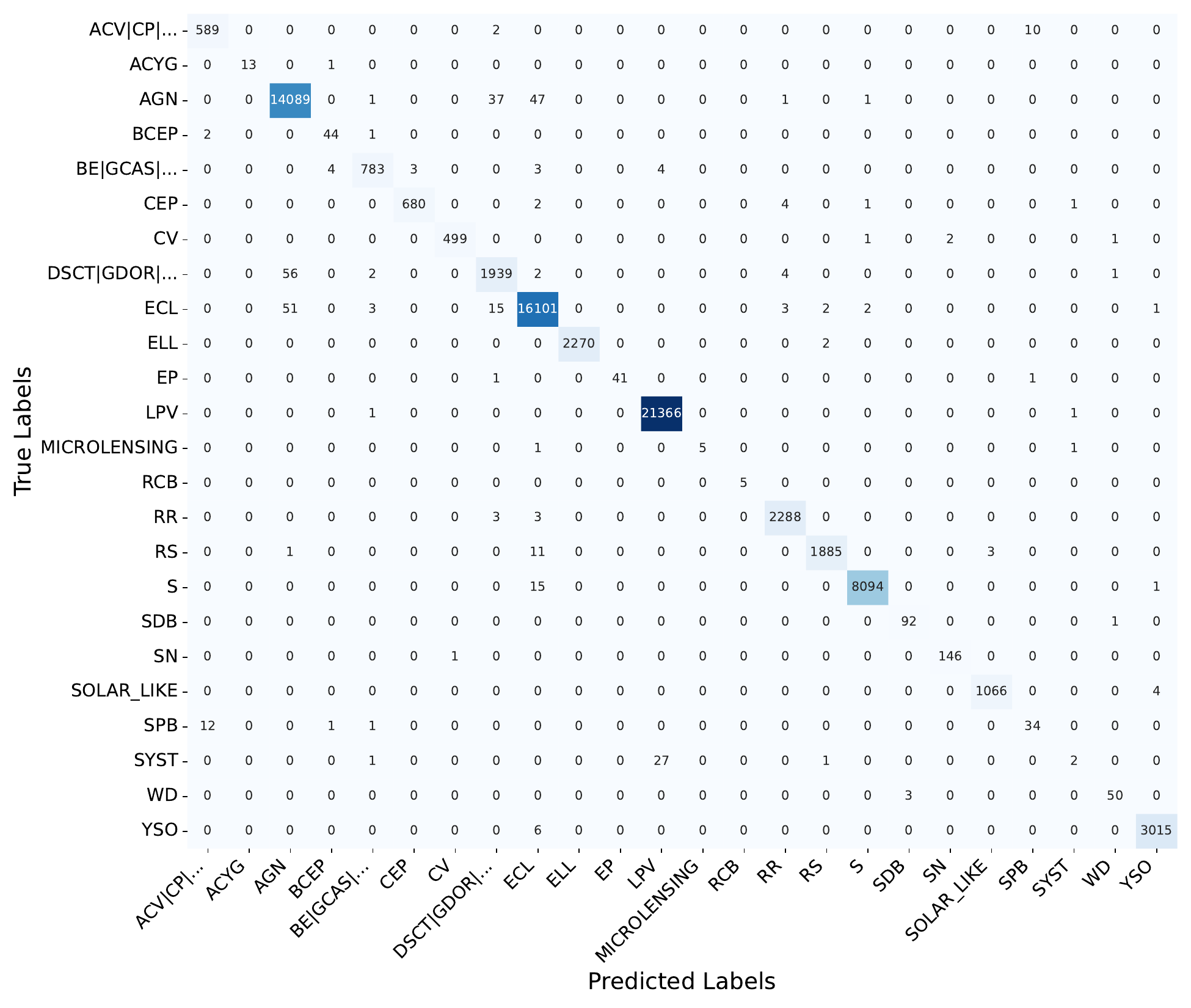}
\caption{Confusion matrix of the StarCLR model on the Gaia fine-tuning task.}
\label{fig:GaiaConfusionMatrix}
\end{figure}

The Gaia dataset was divided into 264,133 samples for training, 37,730 for validation, and 75,470 for testing. The bottom panel of Table~\ref{tab:PerClassAll} summarizes the classification performance on the Gaia test set. The Gaia setting involves a certain degree of out-of-distribution data relative to TESS and ZTF, due to differences in preprocessing. StarCLR achieves a macro $F_1$ score of $92.73\% \pm 0.46\%$ and a micro-$F_1$ score of $99.49\% \pm 0.01\%$. The corresponding confusion matrix is shown in Figure~\ref{fig:GaiaConfusionMatrix}. Overall, the model performs reliably across most major classes, particularly those with sufficient training samples. However, some rare categories remain challenging. For example, the MICROLENSING class attains an $F_1$ score of $86.32\% \pm 5.18\%$, the SPB class reaches $73.13\% \pm 2.16\%$, and the SYST class achieves only $9.53\% \pm 3.01\%$. These reduced performances are largely attributable to limited sample sizes and diverse variability characteristics. Notably, compared with TESS and ZTF, which are dominated by periodic or quasi-periodic variables, Gaia additionally includes non-periodic sources such as AGN and SN.

\section{Discussion} \label{sec:discussion}

This section interprets the observed performance trends and identifies key factors governing the effectiveness of StarCLR. We examine model architectures and training strategies, analyze input embedding design and pooling operations, and study learned representation structure through UMAP visualizations.

\subsection{Model Comparison} \label{subsec:compareModel}

In this subsection, we compare model architectures and training strategies to understand where StarCLR’s performance gains originate. By benchmarking StarCLR against baseline models and ablated variants, we evaluate the effects of architecture, pretraining scheme, and contrastive strategy. In particular, we assess the contribution of representation pretraining and feature usage, investigate the impact of different pretraining datasets on downstream performance, and quantify the benefits of hierarchical contrastive learning to representation quality.

\subsubsection{Architecture and Training Strategy} \label{subsubsec:ohterModels}

To comprehensively evaluate the performance of StarCLR, we conducted comparative experiments on three datasets: TESS, ZTF, and Gaia. We considered three representative configurations: (i) a two-layer LSTM model with a hidden size of 512, which takes the same normalized flux inputs as StarCLR, combined with the corresponding time-aware positional encodings; (ii) a Transformer model trained from scratch, which shares the same architecture as StarCLR but is randomly initialized; and (iii) pretrained StarCLR, where the Transformer backbone was initialized from contrastive pretraining on the TESS dataset and kept frozen during downstream training, with only the classification head optimized. To ensure a fair comparison, all models were trained and evaluated using identical data splits and training protocols. The classification head followed the design illustrated in Figure~\ref{fig:ModelFineTuningArchitecture}. Overall, StarCLR generally outperforms the two comparison models across the three datasets. The improvement is most pronounced on ZTF, where StarCLR achieves a higher macro-$F_1$ of $87.82\%$, compared to $82.15\%$ for the Transformer trained from scratch and $79.51\%$ for the two-layer LSTM. On TESS and Gaia, StarCLR also attains slightly higher macro-$F_1$ scores than both models.

\begin{figure}[ht]
\centering
\includegraphics[width=\linewidth]{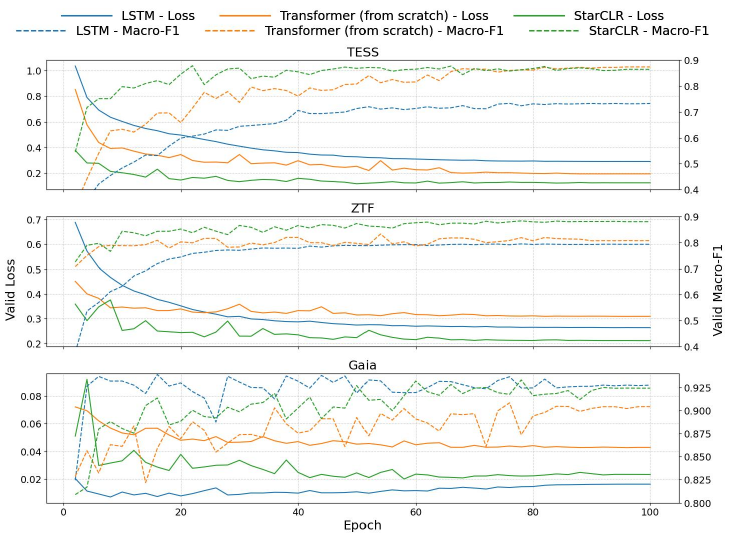}
\caption{Fine-tuning results of three models on the TESS, ZTF, and Gaia datasets. The horizontal axis denotes training epochs; the left vertical axis shows validation loss with solid lines, and the right vertical axis shows validation macro-$F_1$ with dashed lines. From top to bottom are the results on TESS, ZTF, and Gaia validation sets, respectively.}
\label{fig:FineTuningLoss}
\end{figure}

Figure~\ref{fig:FineTuningLoss} shows fine-tuning dynamics on the validation sets of TESS, ZTF, and Gaia. For each run, we select the checkpoint with the highest validation macro-$F_1$ and evaluate it on the corresponding test set to report macro-$F_1$ and micro-$F_1$. Table~\ref{tab:architectureComparison} summarizes the final test results.

On the TESS dataset, the two Transformer-based models achieve substantially higher macro-$F_1$ scores than the LSTM. Specifically, the Transformer trained from scratch reaches $83.91\%$, and StarCLR attains $84.35\%$, whereas the 2-layer LSTM only achieves $68.30\%$. This trend is consistent with the characteristics of TESS light curves: a typical sector spans $\sim 27.4$ days and often contains $\sim 10^3$ cadence points at 30-minute sampling, while 10-minute or 2-minute cadences can produce even longer sequences. For such long and densely sampled sequences, LSTM process observations sequentially and are more prone to losing information over long horizons, which makes it difficult to capture long-range temporal dependencies. In contrast, Transformer-style temporal modeling better preserves global context over long sequences, leading to markedly stronger performance. 

On the ZTF dataset, StarCLR yields the best performance among the three configurations, achieving a macro-$F_1$ of $87.82\%$, outperforming both the Transformer trained from scratch ($82.15\%$) and the LSTM ($79.51\%$). During training, the Transformer trained from scratch reaches a validation macro-$F_1$ of about 0.8 within the first ten epochs, after which progress becomes marginal. This behavior reflects the challenges posed by ZTF light curves, whose sparse sampling often leads to weak or unstable phase relationships between measurements, thereby limiting the effective information content. Under such conditions, the attention mechanism in a randomly initialized Transformer tends to aggregate globally across time, inevitably incorporating low-quality or spurious correlations, which makes periodic patterns difficult to extract and can trap the model in a suboptimal solution. In contrast, the pretrained representations learned by StarCLR provide a more robust temporal prior, alleviating the impact of noisy associations and enabling more effective downstream optimization on ZTF.

On the Gaia dataset, all three models achieve similarly high macro-$F_1$ scores, with StarCLR reaching $92.73\%$, compared to $91.57\% $ for the Transformer trained from scratch and $91.38\%$ for the LSTM. The performance differences among the three models are not statistically significant (t-test, $\alpha=0.05$). This limited separation can be attributed to the Gaia classification scheme, where broad variable-star categories are already well distinguished by physical parameters, providing most of the discriminative power. In this dataset, the temporal representations mainly play a complementary role, and even relatively simple sequence models are sufficient to achieve strong performance. A more detailed analysis of the respective contributions of temporal representations and physical features is presented in Section~\ref{subsubsec:pretrainAblation}.

Although a Transformer trained from scratch could in principle reach comparable performance with extensive hyperparameter tuning, such tuning is computationally expensive and was not exhaustively explored here. In contrast, the pretrained StarCLR backbone provides a strong, stable initialization, allowing downstream optimization to focus on a lightweight MLP classification head. Specifically, the LSTM and scratch Transformer update all parameters for dozens of epochs, whereas StarCLR freezes the backbone and trains only an MLP head after a one-time representation extraction and caching pass. This yields substantially lower training cost and a smaller hyperparameter search space.

\begin{table}[htbp]
\centering
\caption{Comparison of different model architectures on the test sets of the TESS, ZTF, and Gaia datasets. The LSTM and Transformer are trained from scratch, while StarCLR uses a pretrained backbone with frozen parameters.}
\label{tab:architectureComparison}
\begin{tabular}{lccc}
\toprule
Model Configuration & Survey & Macro-$F_1$ (\%) & Micro-$F_1$ (\%) \\
\midrule
\multirow{3}{*}{2-layer LSTM}
 & TESS & $68.30 \pm 0.17$ & $91.27 \pm 0.20$ \\
 & ZTF  & $79.51 \pm 0.18$ & $91.07 \pm 0.08$ \\
 & Gaia & $91.38 \pm 0.56$ & $99.81 \pm 0.01$ \\
\midrule
\multirow{3}{*}{Transformer (from scratch)}
 & TESS & $83.91 \pm 0.79$ & $94.94 \pm 0.11$ \\
 & ZTF  & $82.15 \pm 0.63$ & $89.00 \pm 0.22$ \\
 & Gaia & $91.57 \pm 0.58$ & $98.63 \pm 0.02$ \\
\midrule
\multirow{3}{*}{StarCLR}
 & TESS & $84.35 \pm 1.32$ & $96.46 \pm 0.27$ \\
 & ZTF  & $87.82 \pm 0.29$ & $92.83 \pm 0.06$ \\
 & Gaia & $92.73 \pm 0.46$ & $99.49 \pm 0.01$ \\
\bottomrule
\end{tabular}
\end{table}

\subsubsection{Effect of Pretraining and Feature Ablation} \label{subsubsec:pretrainAblation}

To systematically evaluate the effectiveness of the representations learned by StarCLR and to disentangle the respective contributions of pretrained representations, astrophysical features, and classifier capacity to the final performance, we conduct a series of comparative and ablation experiments on the TESS, ZTF, and Gaia datasets. Overall, the results indicate that contrastive pretraining enables the model to capture useful temporal structures in light curves, although the extent of performance gains depends strongly on the characteristics of the target dataset, such as sampling patterns and classification granularity.

In this subsection, we adopt both a 1-layer MLP and a 4-layer MLP as classification heads, enabling controlled ablations over classifier capacity and input configurations. In all settings, the Transformer backbone is kept frozen so that the analysis primarily reflects representation quality. The main results are summarized in Table~\ref{tab:pretrainFeatureAblation}. 

We first compare a contrastively pretrained StarCLR backbone with a randomly initialized Transformer under the same classification head to quantify the impact of pretraining. With a 1-layer MLP, contrastive pretraining improves macro-$F_1$ by approximately 32.9, 10.7, and 3.6 percentage points on TESS, ZTF, and Gaia, respectively; when using a 4-layer MLP, the corresponding improvements increase to about 40.2, 11.4, and 6.4 percentage points. These results suggest that pretraining improves representation quality, although the magnitude of improvement is notably smaller on Gaia. Meanwhile, using backbone representations only, replacing the 1-layer head with a 4-layer MLP further improves macro-$F_1$ by approximately 12.1, 6.8, and 6.9 percentage points on TESS, ZTF, and Gaia, respectively, suggesting that a higher-capacity and more nonlinear classifier can better leverage the features extracted by the pretrained backbone, thereby yielding larger performance gains.

We then quantify the relative contributions of backbone representations and astrophysical features. Under the 1-layer MLP setting, the macro-$F_1$ differences between the pretrained backbone model and the feature-only model are approximately 2.3, 19.1, and 53.1 percentage points on TESS, ZTF, and Gaia, respectively. When the classification head is increased to a 4-layer MLP, these gaps further expand to about 21.5, 34.2, and 53.3 percentage points, with the feature-only model consistently achieving higher scores across the three datasets. This result suggests that, compared with relying solely on light-curve representations, explicitly incorporating astrophysical features provides stronger discriminative information under the current task settings. One reason is that variable-star classification typically depends on both physical quantities—such as period, luminosity, and color—and morphological characteristics of the light curve. Many traditional classification criteria and survey catalog pipelines are more directly driven by period and physical parameters; therefore, feature-based models naturally contain stronger prior discriminative signals. In contrast, the pretrained backbone learns primarily from normalized light curves and captures morphological and relative temporal structures. Although it can implicitly encode some period-related information, it lacks quantities that depend on distance and extinction corrections, which may lead to larger performance gaps on certain datasets. The variation in the magnitude of these differences across datasets is also closely related to classification granularity. TESS adopts the most fine-grained subclass taxonomy, followed by ZTF, while Gaia uses a relatively coarser classification scheme. As classification becomes finer, distinguishing subclasses increasingly relies on subtle morphological differences in light curves. Consequently, on TESS, the advantage of feature-only models is relatively constrained and temporal representations become more important; on Gaia, where the taxonomy is coarser, astrophysical features are likely to provide strong discriminative power. Finally, we also note that distribution mismatch between the pretraining data and downstream datasets can further affect transfer performance (see Section~\ref{subsubsec:noHierarchicalPretrain}).

\begin{table}[htbp]
\centering
\caption{Effect of pretraining and feature ablation under different classifier settings.}
\label{tab:pretrainFeatureAblation}
\begin{tabular}{lccc}
\toprule
Setting & Survey & Macro-$F_1$ (\%) & Micro-$F_1$ (\%) \\
\midrule
\multirow{3}{*}{Random Transformer Backbone + 1-layer MLP}
 & TESS & $10.35 \pm 0.08$ & $57.17 \pm 0.86$ \\
 & ZTF  & $29.28 \pm 0.74$ & $72.10 \pm 0.40$ \\
 & Gaia & $25.99 \pm 0.29$ & $57.96 \pm 0.90$ \\
\midrule
\multirow{3}{*}{Random Transformer Backbone + 4-layer MLP}
 & TESS & $15.16 \pm 0.31$ & $66.98 \pm 0.52$ \\
 & ZTF  & $35.35 \pm 0.07$ & $74.36 \pm 0.06$ \\
 & Gaia & $30.07 \pm 0.38$ & $62.33 \pm 0.34$ \\
\midrule
\multirow{3}{*}{StarCLR Backbone + 1-layer MLP}
 & TESS & $43.29 \pm 0.62$ & $83.66 \pm 0.08$ \\
 & ZTF  & $39.97 \pm 0.41$ & $76.10 \pm 0.23$ \\
 & Gaia & $29.54 \pm 0.25$ & $64.57 \pm 0.15$ \\
\midrule
\multirow{3}{*}{StarCLR Backbone + 4-layer MLP}
 & TESS & $55.37 \pm 0.55$ & $88.03 \pm 0.18$ \\
 & ZTF  & $46.72 \pm 0.06$ & $78.89 \pm 0.03$ \\
 & Gaia & $36.42 \pm 0.19$ & $74.28 \pm 0.08$ \\
\midrule
\multirow{3}{*}{Astrophysical Features Only + 1-layer MLP}
 & TESS & $45.58 \pm 0.24$ & $79.79 \pm 0.15$ \\
 & ZTF  & $59.09 \pm 0.26$ & $78.08 \pm 0.07$ \\
 & Gaia & $82.60 \pm 0.84$ & $95.71 \pm 0.05$ \\
\midrule
\multirow{3}{*}{Astrophysical Features Only + 4-layer MLP}
 & TESS & $76.87 \pm 0.57$ & $92.87 \pm 0.10$ \\
 & ZTF  & $80.94 \pm 0.03$ & $88.79 \pm 0.07$ \\
 & Gaia & $89.67 \pm 0.31$ & $98.31 \pm 0.09$ \\
\midrule
\multirow{3}{*}{StarCLR Backbone + feature + 1-layer MLP}
 & TESS & $73.33 \pm 0.18$ & $91.91 \pm 0.02$ \\
 & ZTF  & $74.55 \pm 0.94$ & $86.06 \pm 0.36$ \\
 & Gaia & $87.86 \pm 0.73$ & $98.53 \pm 0.04$ \\
\midrule
\multirow{3}{*}{StarCLR Backbone + feature + 4-layer MLP}
 & TESS & $84.35 \pm 1.32$ & $96.46 \pm 0.27$ \\
 & ZTF  & $87.82 \pm 0.29$ & $92.83 \pm 0.06$ \\
 & Gaia & $92.73 \pm 0.46$ & $99.49 \pm 0.01$ \\
\bottomrule
\end{tabular}
\end{table}

We next examine models that rely on astrophysical features. When using astrophysical features only, increasing the classification head from a 1-layer to a 4-layer MLP improves macro-$F_1$ by approximately 31.3, 21.9, and 7.1 percentage points on TESS, ZTF, and Gaia, respectively. We further compare the performance gain of combining astrophysical features with light-curve representations over using astrophysical features alone. With a 1-layer MLP, the fused model increases macro-$F_1$ by about 27.8, 15.5, and 5.3 percentage points on TESS, ZTF, and Gaia, respectively; with a 4-layer MLP, the corresponding gains decrease to approximately 7.5, 6.9, and 3.1 percentage points. These results are consistent with the analysis above: as the classification becomes finer-grained and the task relies more on morphological discrimination, the contribution of light-curve representations becomes more pronounced. Accordingly, the largest improvement is observed on TESS, followed by ZTF, and then Gaia. Meanwhile, we also examine the effect of classifier capacity within the fused setting. Replacing the 1-layer head with a 4-layer MLP further improves macro-$F_1$ by approximately 11.0, 13.3, and 4.9 percentage points on TESS, ZTF, and Gaia, respectively. Because the fused model already achieves strong performance with a 1-layer MLP, increasing classifier capacity yields smaller additional gains and the improvement does not scale linearly.

Taken together, the results suggest that temporal representations learned by StarCLR capture useful information from light curves and provide measurable gains beyond astrophysical features alone. For Gaia, however, the results indicate that the overall performance is driven primarily by astrophysical features rather than by the pretrained temporal backbone. This is already evident from Table~\ref{tab:pretrainFeatureAblation}, where the backbone-only setting yields substantially lower performance than the feature-only model, and the gap remains large even when increasing classifier capacity. To further clarify which classes dominate this behavior, we examine the main contributors in the Gaia test set. AGN, LPV, and ECL together account for about 68.5\% of the total sample, and their $F_1$ scores improve by approximately 30.1, 35.7, and 12.1 percentage points, respectively, after incorporating astrophysical features. These improvements are consistent with their physical properties: AGN is already distinguishable from periodic variables through its non-periodic variability, while magnitude- and parallax-related information further enhance this separation; LPV becomes much easier to identify once period-related information is included; and ECL also benefits from the addition of physical features.

\subsubsection{Comparison of Different Pretraining Datasets} \label{subsubsec:otherPretrain}

To examine how pretraining data characteristics affect downstream classification, we compare models pretrained on two surveys with distinct observation regimes (TESS and ZTF). To isolate representation quality, all experiments use a frozen-backbone setting and a 1-layer MLP classifier.

Quantitative results are summarized in Table~\ref{tab:crossSurveyPretrain}. We find that contrastive pretraining consistently enables StarCLR to learn temporal representations, substantially outperforming a randomly initialized Transformer. Meanwhile, the observational characteristics of the pretraining data have a measurable impact on downstream performance. Using TESS pretraining as the reference, pretraining on the more sparsely and irregularly sampled ZTF data reduces the macro-$F_1$ on TESS by approximately 12.7 percentage points. In contrast, the same pretraining improves performance on sparsely sampled datasets, increasing macro-$F_1$ by approximately 8.8 percentage points on ZTF and by approximately 8.6 percentage points on Gaia. These results suggest that the temporal structures emphasized during pretraining bias the learned representations toward the characteristics of the pretraining data: dense-cadence TESS pretraining performs best on TESS itself, whereas ZTF pretraining is more favorable for sparsely and irregularly sampled downstream datasets. Given that the TESS classification task involves a larger number of classes and more detailed subclass distinctions than those of Gaia and ZTF in our setting, we adopt the StarCLR model pretrained on TESS in this work.

\begin{table}[htbp]
\centering
\caption{Effect of different pretraining data on downstream classification performance.}
\label{tab:crossSurveyPretrain}
\begin{tabular}{lccc}
\toprule
Setting & Survey & Macro-$F_1$ (\%) & Micro-$F_1$ (\%) \\
\midrule
\multirow{3}{*}{Random Transformer Backbone + 1-layer MLP}
 & TESS & $10.35 \pm 0.08$ & $57.17 \pm 0.86$ \\
 & ZTF  & $29.28 \pm 0.74$ & $72.10 \pm 0.40$ \\
 & Gaia & $25.99 \pm 0.29$ & $57.96 \pm 0.90$ \\
\midrule
\multirow{3}{*}{StarCLR Backbone (pretrained on TESS) + 1-layer MLP}
 & TESS & $43.29 \pm 0.62$ & $83.66 \pm 0.08$ \\
 & ZTF  & $39.97 \pm 0.41$ & $76.10 \pm 0.23$ \\
 & Gaia & $29.54 \pm 0.25$ & $64.57 \pm 0.15$ \\
\midrule
\multirow{3}{*}{StarCLR Backbone (pretrained on ZTF) + 1-layer MLP}
 & TESS & $30.55 \pm 0.43$ & $77.97 \pm 0.24$ \\
 & ZTF  & $48.76 \pm 0.28$ & $79.69 \pm 0.34$ \\
 & Gaia & $38.10 \pm 0.38$ & $75.20 \pm 0.22$ \\
\bottomrule
\end{tabular}
\end{table}

\subsubsection{Effect of Hierarchical Contrastive Learning} \label{subsubsec:noHierarchicalPretrain}

As described in Section~\ref{subsubsec:hierarchicalLoss}, we adopt hierarchical contrastive learning during pretraining. To evaluate its effectiveness, we compare it with a non-hierarchical scheme. In both settings, the pretrained backbone is frozen and a 1-layer MLP is used as the classification head to minimize the influence of classifier capacity.

Table~\ref{tab:hierarchicalContrastiveAblation} summarizes the comparison between hierarchical and non-hierarchical contrastive pretraining. Hierarchical pretraining improves macro-$F_1$ by approximately 7.5 percentage points on TESS, 4.2 percentage points on ZTF, and 2.7 percentage points on Gaia. These gains suggest that incorporating multi-scale temporal structure during pretraining leads to more discriminative and robust representations.

\begin{table}[htbp]
\centering
\caption{Effect of hierarchical contrastive learning.}
\label{tab:hierarchicalContrastiveAblation}
\begin{tabular}{lccc}
\toprule
Pretraining Strategy & Survey & Macro-$F_1$ (\%) & Micro-$F_1$ (\%) \\
\midrule
\multirow{3}{*}{Non-hierarchical Contrastive Learning}
 & TESS & $35.82 \pm 0.12$ & $80.92 \pm 0.11$ \\
 & ZTF  & $35.76 \pm 0.40$ & $75.41 \pm 0.19$ \\
 & Gaia & $26.87 \pm 0.18$ & $61.14 \pm 0.08$ \\
\midrule
\multirow{3}{*}{Hierarchical Contrastive Learning}
 & TESS & $43.29 \pm 0.62$ & $83.66 \pm 0.08$ \\
 & ZTF  & $39.97 \pm 0.41$ & $76.10 \pm 0.23$ \\
 & Gaia & $29.54 \pm 0.25$ & $64.57 \pm 0.15$ \\
\bottomrule
\end{tabular}
\end{table}

\subsection{Input Embedding Design} \label{subsec:embeddingAnalysis}

To analyze how input embedding design affects temporal representation learning, we compare three methods for encoding time and flux. The first two use time-aware positional encoding based on real observation times, explicitly injecting temporal interval information into the Transformer. They differ only in whether the angular frequency $\omega_i$ includes a $2\pi$ scaling factor; formulations are provided in Section~\ref{subsec:PE}. The third normalizes time and flux separately to $[0,1]$, concatenates them, and projects the result with a linear layer to form a joint time--flux embedding. Overall, time-aware positional encoding with $2\pi$ scaling yields the best performance.

For comparing these three methods, the pretrained backbone is kept frozen and a 1-layer MLP is used as the classification head. Quantitative results are summarized in Table~\ref{tab:embeddingDesign}. For the joint embedding, time normalization is defined as
\begin{equation}
t' = \frac{t - \min(t)}{\max(t) - \min(t)}
\end{equation}

From a representation perspective, the primary difference among these methods lies in how temporal information is extracted and combined with flux. The joint embedding first applies min--max scaling to time and flux independently, then concatenates them and uses a linear projection to obtain a high-dimensional embedding. In contrast, the time-aware positional encoding applies sinusoidal transformations to observation time, introducing an inherent nonlinearity in the mapping from time to the embedding space.

The impact of embedding design is most pronounced on the TESS dataset. TESS contains sequences with multiple short cadences of 30~min, 10~min, and 2~min over a baseline of about 27.4~days. Under the joint embedding, compressing time into the range $[0,1]$ can make adjacent timestamps differ only marginally on the normalized scale, which limits the ability of a linear projection to represent fine-grained temporal structure. By contrast, sinusoidal positional encoding performs a nonlinear transformation before the observation time is fed into the model. This nonlinear mapping expands small time differences into more distinctive patterns in the embedding space, enabling richer temporal representations. As a result, on densely sampled data where short time intervals play an important role, such as TESS, time-aware positional encoding without the $2\pi$ scaling produces substantially more informative temporal representations, improving macro-$F_1$ by approximately 15.1 percentage points compared to the joint time--flux embedding. For ZTF and Gaia, where sampling is sparser and characteristic time intervals are longer, the information loss introduced by normalization is less severe, and the linear joint embedding is often sufficient to convey the dominant temporal cues. Accordingly, switching from the joint embedding to time-aware encoding changes macro-$F_1$ by only a few percentage points on ZTF and Gaia, indicating much weaker sensitivity than on TESS.

A direct comparison between the two time-aware positional encoding further shows that introducing the $2\pi$ scaling maps the projected time at each scale to an angular variable, making the sinusoidal encoding more explicitly capture phase-like periodic components across multiple characteristic timescales. Since astronomical light curves commonly exhibit periodic or quasi-periodic structures, this phase-sensitive time encoding is more likely to yield additional gains on densely sampled surveys such as TESS, while providing more limited benefits under sparse sampling. Empirically, the $2\pi$ scaling brings an additional improvement of approximately 6.6 percentage points on TESS, but only about 0.8 percentage points on ZTF and Gaia.

\begin{table}[htbp]
\centering
\caption{Effect of different input embedding designs.}
\label{tab:embeddingDesign}
\begin{tabular}{lccc}
\toprule
Embedding Design & Survey & Macro-$F_1$ (\%) & Micro-$F_1$ (\%) \\
\midrule
\multirow{3}{*}{Joint Time--Flux Embedding}
 & TESS & $21.57 \pm 0.62$ & $72.78 \pm 0.88$ \\
 & ZTF  & $41.32 \pm 0.99$ & $76.42 \pm 0.54$ \\
 & Gaia & $26.98 \pm 0.47$ & $63.94 \pm 0.21$ \\
\midrule
\multirow{3}{*}{Time-aware Positional Encoding (w/o $2\pi$)}
 & TESS & $36.65 \pm 1.74$ & $80.19 \pm 0.23$ \\
 & ZTF  & $39.21 \pm 0.27$ & $75.62 \pm 0.22$ \\
 & Gaia & $28.76 \pm 0.15$ & $64.65 \pm 0.18$ \\
\midrule
\multirow{3}{*}{Time-aware Positional Encoding (w/ $2\pi$)}
 & TESS & $43.29 \pm 0.62$ & $83.66 \pm 0.08$ \\
 & ZTF  & $39.97 \pm 0.41$ & $76.10 \pm 0.23$ \\
 & Gaia & $29.54 \pm 0.25$ & $64.57 \pm 0.15$ \\
\bottomrule
\end{tabular}
\end{table}

\subsection{Pooling Analysis} \label{subsec:poolingAnalysis}

In this subsection, we investigate pooling strategies under a frozen-backbone setting with a 1-layer MLP classification head. As described in Section~\ref{subsubsec:hierarchicalLoss}, we consider max and mean pooling. Each is applied during contrastive pretraining and then combined with either method during fine-tuning. Evaluations on TESS, ZTF, and Gaia show that pooling choice during pretraining has only a marginal effect on downstream performance, whereas final results are more sensitive to pooling used in fine-tuning.

The quantitative results are summarized in Table~\ref{tab:poolingDesign}, where the notation ``Pre'' and ``FT'' indicate the pooling strategy adopted in the pretraining and fine-tuning stages, respectively. Overall, performance differences associated with different pretraining pooling strategies remain small. In particular, when mean pooling is used during fine-tuning, switching the pretraining pooling between max and mean leads to little change on TESS and Gaia, with only a modest variation of about 2.5 percentage points on ZTF. In contrast, the pooling strategy employed during fine-tuning has a more pronounced impact on the final classification performance. When max pooling is used during pretraining, on TESS the difference between fine-tuning pooling choices remains minor, whereas on ZTF and Gaia the effect is much clearer: fine-tuning with mean pooling improves macro-$F_1$ by approximately 4.6 percentage points on ZTF and 4.9 percentage points on Gaia compared to max pooling during fine-tuning. A similar trend is observed when mean pooling is used during pretraining. This trend is consistent across different pretraining pooling settings, suggesting that mean pooling during fine-tuning often yields more favorable performance.

From a modeling perspective, mean pooling during fine-tuning can better smooth noise and extreme values in irregularly sampled light curves, leading to more robust global sequence representations and improved generalization. Based on this observation, we consistently adopt mean pooling as the aggregation method during fine-tuning in our comparisons with the Transformer trained from scratch. Meanwhile, during pretraining, we retain max pooling as the default strategy to align with the design motivation of hierarchical contrastive learning, which aims to emphasize the most salient variations across sub-sequences.

\begin{table}[htbp]
\centering
\caption{Effect of different pooling strategies in pretraining and fine-tuning stages.}
\label{tab:poolingDesign}
\begin{tabular}{lccc}
\toprule
Pooling Design & Survey & Macro-$F_1$ (\%) & Micro-$F_1$ (\%) \\
\midrule
\multirow{3}{*}{MaxPool (Pre) + MaxPool (FT)}
 & TESS & $44.74 \pm 0.69$ & $82.58 \pm 0.38$ \\
 & ZTF  & $35.36 \pm 1.18$ & $72.91 \pm 1.40$ \\
 & Gaia & $24.74 \pm 0.32$ & $57.19 \pm 1.03$ \\
\midrule
\multirow{3}{*}{MaxPool (Pre) + MeanPool (FT)}
 & TESS & $43.29 \pm 0.62$ & $83.66 \pm 0.08$ \\
 & ZTF  & $39.97 \pm 0.41$ & $76.10 \pm 0.23$ \\
 & Gaia & $29.65 \pm 0.28$ & $64.73 \pm 0.14$ \\
\midrule
\multirow{3}{*}{MeanPool (Pre) + MaxPool (FT)}
 & TESS & $42.40 \pm 1.24$ & $84.43 \pm 0.30$ \\
 & ZTF  & $30.73 \pm 0.44$ & $71.44 \pm 1.08$ \\
 & Gaia & $23.79 \pm 0.56$ & $53.46 \pm 1.64$ \\
\midrule
\multirow{3}{*}{MeanPool (Pre) + MeanPool (FT)}
 & TESS & $43.50 \pm 0.06$ & $85.38 \pm 0.08$ \\
 & ZTF  & $37.44 \pm 0.08$ & $75.74 \pm 0.18$ \\
 & Gaia & $29.91 \pm 0.08$ & $64.09 \pm 0.05$ \\
\bottomrule
\end{tabular}
\end{table}

\subsection{UMAP analysis} \label{subsec:UMAP}

In this subsection, we employ Uniform Manifold Approximation and Projection \citep[UMAP;][]{UMAP} to visualize features. UMAP is a nonlinear manifold-learning dimensionality reduction method that maps high-dimensional data into a low-dimensional space for visualization and structural exploration. Here, we reduce features to two dimensions. Consistent with fine-tuning, we first apply mean pooling along the temporal dimension and then concatenate supplementary features to construct the final high-dimensional vector. The vectors are projected onto a two-dimensional plane for each survey to examine clustering behavior in feature space. Overall, the UMAP projections show that StarCLR representations effectively distinguish most major variable-star classes, while overlaps remain between certain subclasses and some categories appear scattered.

\begin{figure}[ht]
\centering
\includegraphics[width=\linewidth]{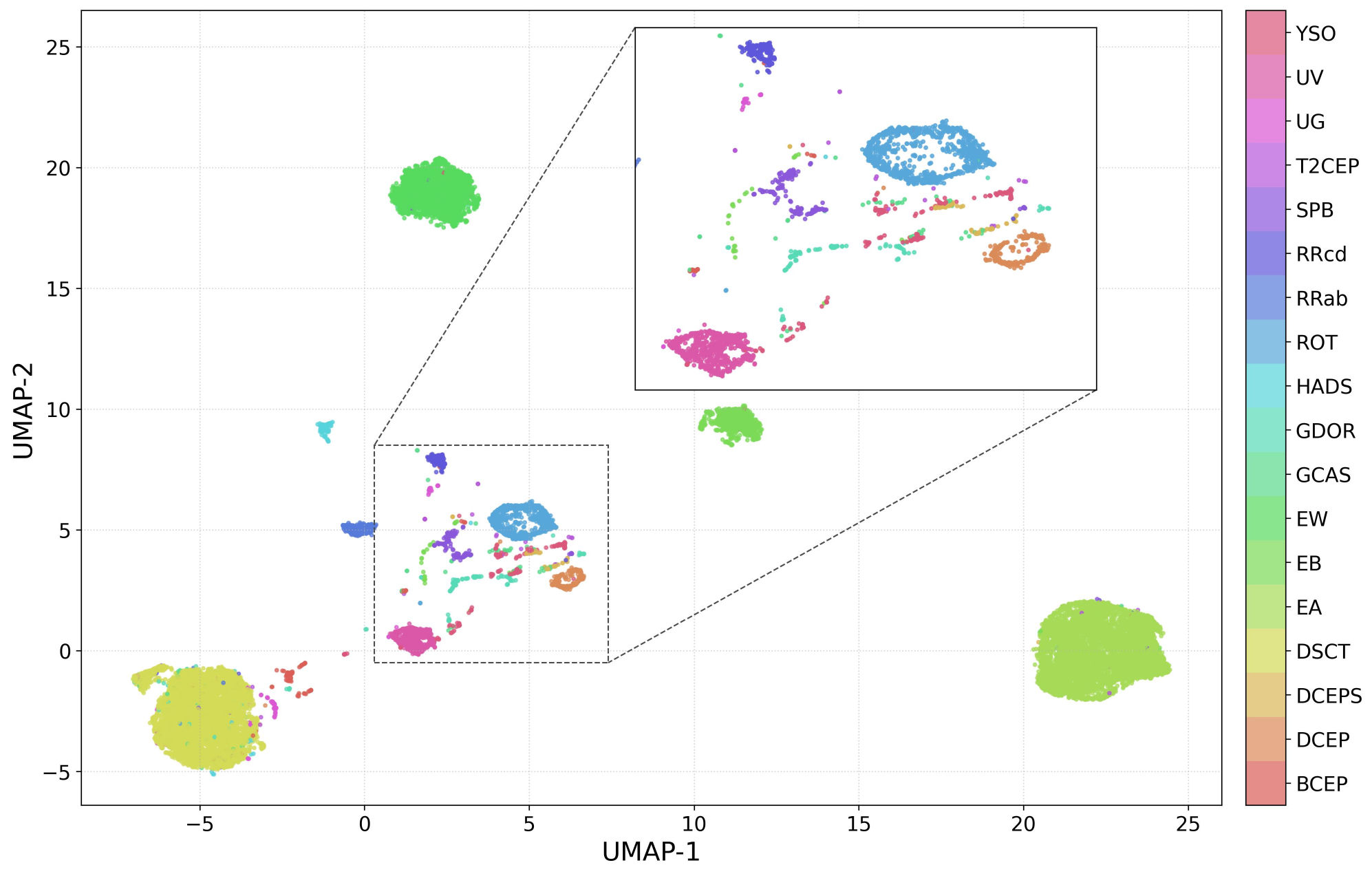}
\caption{UMAP projection of the TESS dataset, showing the global distribution and a locally magnified region. UMAP-1 and UMAP-2 denote the two embedding dimensions obtained after dimensionality reduction.}
\label{fig:UMAP_TESS}
\end{figure}

Figure~\ref{fig:UMAP_TESS} shows the visualization of the TESS data. The UMAP-1 and UMAP-2 axes do not correspond to physical quantities; instead, they represent the two coordinates produced by UMAP when projecting high-dimensional feature vectors into a two-dimensional embedding space. At a global level, the learned representations exhibit clear clustering for most major variable star types, indicating strong feature discriminability. For example, the EA, EB, EW, and DSCT classes occupy relatively distinct regions in the low-dimensional space. However, DSCT partially overlaps with other categories, consistent with the confusion matrix, where DSCT tends to absorb misclassified EA and EW samples. For Cepheids, the fundamental-mode (DCEP), first-overtone (DCEPS), and Type~II Cepheids (T2CEP) show partial overlap in the projection space, indicating the limitations of the learned representations in capturing fine-grained differences among these subclasses, which is also reflected in the classification results.

\begin{figure}[ht]
\centering
\includegraphics[width=\linewidth]{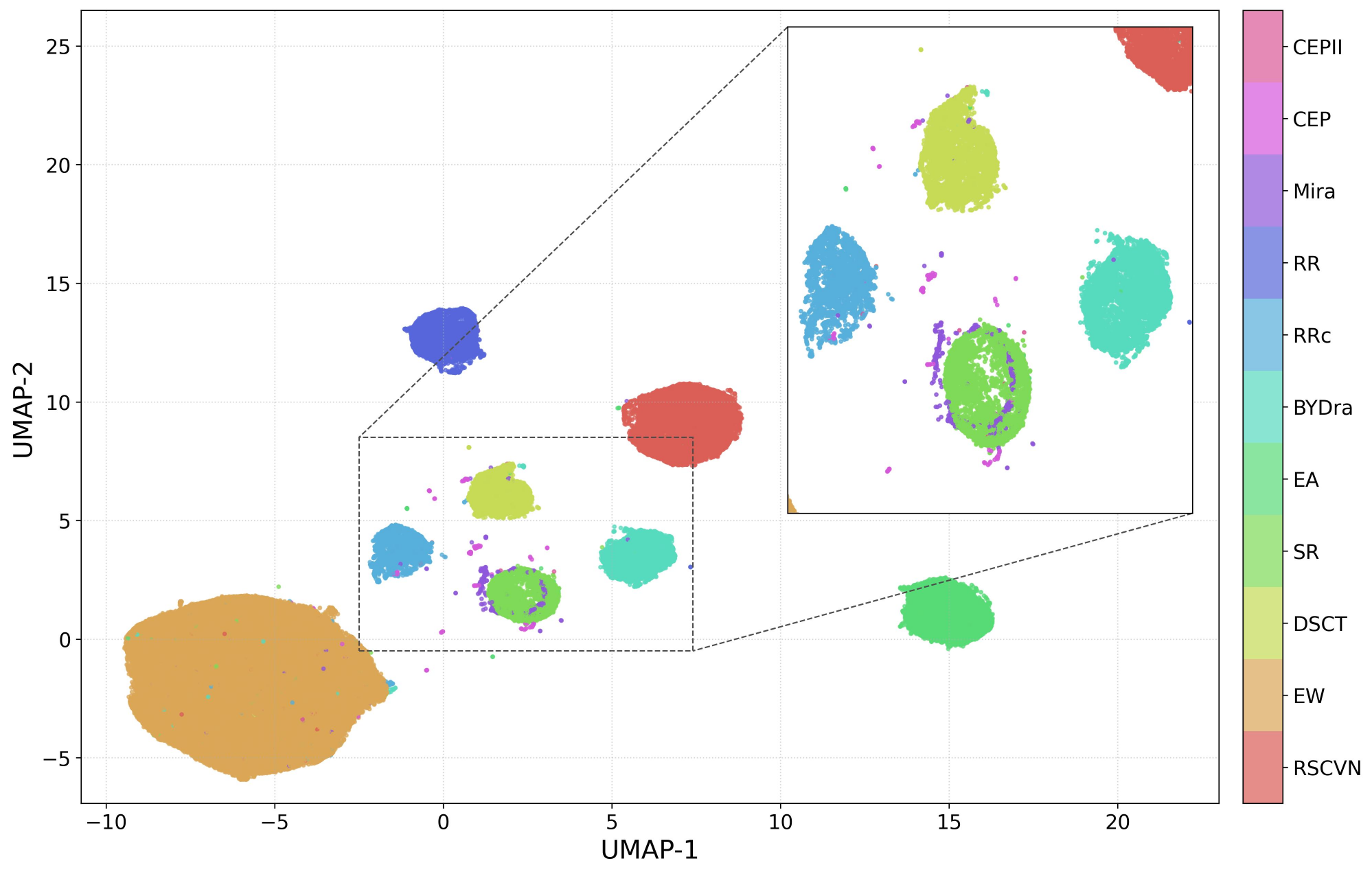}
\caption{UMAP projection of the ZTF dataset, showing the global distribution and a locally magnified region.}
\label{fig:UMAP_ZTF}
\end{figure}

Figure~\ref{fig:UMAP_ZTF} presents the UMAP projection of ZTF data. Owing to the larger sample size, variable star classes exhibit clearer clustering boundaries in the embedding space. Nevertheless, typical mixing phenomena are observed. For instance, SR and Mira stars show overlapping distributions, consistent with their close relationship in astrophysical classification, where the two are primarily distinguished by amplitude. However, amplitude differences may not be fully preserved in the two-dimensional embedding, leading to overlap in the projection. It is worth noting that although SR and Mira overlap in the projection space, both maintain relatively high $F_1$-scores in the confusion matrix, indicating that the overlap is primarily due to information loss during dimensionality reduction rather than a deficiency of the classifier. Another example is the EW class, where some contamination from multiple variable star types can be observed, possibly due to poor light curve quality or errors in the original labels. In addition, the Cepheid class does not form compact clusters in the ZTF UMAP projection. This behavior is likely related to the limited number of Cepheid samples in the ZTF dataset, which account for only about $0.2\%$ of the total data, resulting in less effective dimensionality reduction for this class.

\begin{figure}[ht]
\centering
\includegraphics[width=\linewidth]{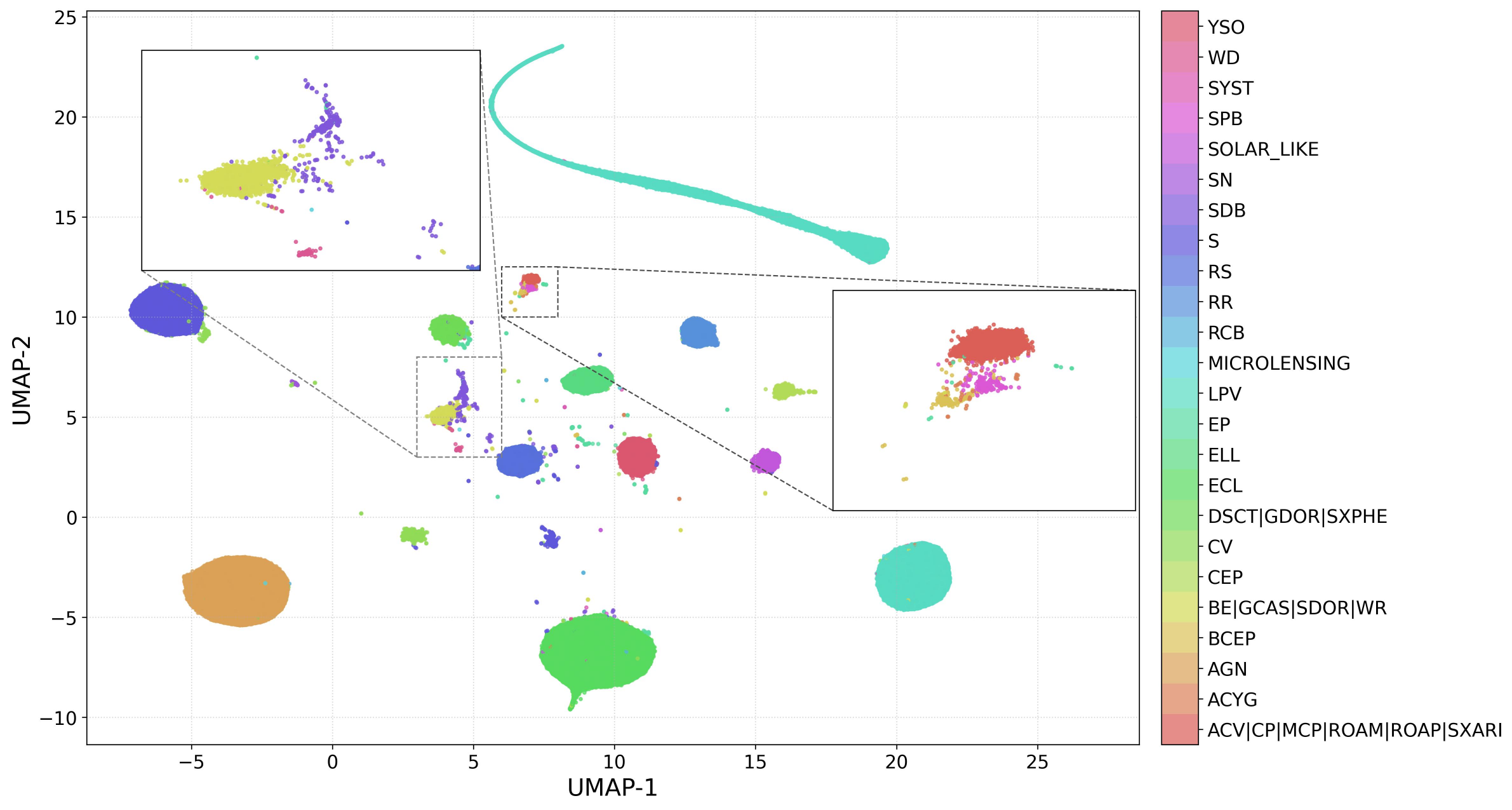}
\caption{UMAP projection of the Gaia dataset, showing the global distribution and a locally magnified region.}
\label{fig:UMAP_Gaia}
\end{figure}

Figure~\ref{fig:UMAP_Gaia} shows the UMAP projection of Gaia data, where scattering phenomena are more pronounced. First, the EP class fails to form a clear cluster due to the small number of samples. Diffuse distributions can also be observed near the center of the figure for some SN and ECL samples, although these classes still form distinct clusters in other regions. This may be attributed to noise: on the one hand, the Transformer inevitably introduces redundant or unstable features during representation learning; on the other hand, the preprocessing strategy may amplify noise. For example, in Gaia data, a large number of missing values (NaNs) are replaced with placeholders, which may introduce additional bias. Nevertheless, Gaia still yields the best classification performance during fine-tuning, suggesting that the MLP classifier can suppress noise to some extent through parameter learning. In UMAP projections, however, high-dimensional information is compressed into two dimensions, inevitably discarding some discriminative features while amplifying residual noise or unstable components, which appear more prominently as scattered points. Furthermore, the LPV class exhibits two distinct clusters located around coordinates (19, 14) and (21, –3), likely corresponding to SR, Mira, and OGLE Small-Amplitude Red Giants, which are separated in the low-dimensional embedding space.

In summary, the UMAP visualizations reveal that the learned representation can effectively separate the major classes of variable stars in feature space, while overlaps and scattering remain for certain subclasses. These results highlight both the challenges in extracting fine-grained features and the influence of data preprocessing and noise on downstream performance.

\section{Conclusion} \label{sec:conclusion}

This work introduces StarCLR, a Transformer-based contrastive pretraining framework for large-scale representation learning of astronomical light curves. Unlike traditional methods that rely on interpolation or regular sampling, StarCLR directly handle irregularly sampled time-series data and construct positive pairs from partially overlapping subcurves, encouraging the model to learn robust temporal representations. Benefiting from the flexibility of the Transformer architecture and the time-aware positional encoding, StarCLR naturally supports variable-length and irregularly sampled sequences, making it applicable to diverse observational modes across different surveys.

During the pretraining stage, we perform contrastive learning on 764,986 TESS light curves from the first 69 sectors, which are split into training, validation, and test sets with a ratio of 7:1:2. The model converges to a test loss of 0.02136. In the fine-tuning stage, we evaluate the pretrained model for variable star classification on labeled datasets from three surveys, including 19,796 samples from TESS, 242,169 samples from ZTF, and 377,333 samples from Gaia. On the test sets, the pretrained model achieves macro-$F_1$ scores of $84.35\% \pm 1.32\%$, $87.82\% \pm 0.29\%$, and $92.73\% \pm 0.46\%$ on the TESS, ZTF, and Gaia datasets, respectively, together with corresponding micro-$F_1$ scores of $96.46\% \pm 0.27\%$, $92.83\% \pm 0.06\%$, and $99.49\% \pm 0.01\%$. Compared with LSTM models and Transformers trained from scratch, the pretrained Transformer demonstrates superior performance on TESS and ZTF, with the most pronounced gains on the sparsely sampled ZTF light curves. For Gaia, although the evaluation involves a certain degree of out-of-distribution data, strong overall performance is still observed. Our ablation results indicate that this is driven primarily by the discriminative power of Gaia astrophysical features. Notably, fine-tuning only the classification head is sufficient to effectively leverage the pretrained backbone representations, substantially reducing training cost and improving flexibility for practical deployment.

We further investigate the effectiveness of the pretrained representations using linear-probe evaluations. By separately applying linear probes to the Transformer backbone representations and to astrophysical features, our ablation results show that performance on downstream tasks is largely driven by astrophysical features. Nevertheless, even in the Gaia setting, where feature-based separability is likely strong, incorporating temporal representations extracted by the pretrained backbone still provides additional performance gains. To assess the effectiveness of StarCLR under different observational characteristics, we conduct additional pretraining experiments on TESS and ZTF data separately and evaluate the resulting models on the TESS, ZTF, and Gaia datasets. In both cases, contrastive pretraining provides consistent downstream benefits. Moreover, pretraining on densely sampled TESS data yields stronger performance on densely sampled datasets, while pretraining on sparsely sampled ZTF data leads to more favorable results on sparsely sampled datasets such as ZTF and Gaia. This suggests that aligning the observational characteristics of the pretraining data with those of the target dataset can improve downstream performance. In addition, ablation experiments comparing standard and hierarchical contrastive learning strategies indicate that the hierarchical contrastive objective produces representations that are better aligned with downstream classification tasks, further supporting its effectiveness within the StarCLR framework.

We also compare different embedding and pooling designs within the StarCLR framework. The results show that encoding temporal information with nonlinear trigonometric positional embeddings leads to more effective downstream performance, with the variant that incorporates a $2\pi$ scaling in the angular frequency consistently outperforming its counterpart. Regarding pooling strategies, ablation experiments indicate that the choice of pooling has a limited impact during the pretraining stage, whereas it plays a more important role in downstream classification. In particular, mean pooling in the classification head yields more stable and favorable performance across datasets. In addition, UMAP visualizations provide qualitative evidence that the learned representations effectively separate most major classes of variable stars, while also revealing residual overlaps among closely related subclasses and more scattered distributions for minority classes.

Future work will explore joint pretraining and mixed fine-tuning across TESS, ZTF, and Gaia, and extend classification to finer-grained taxonomies with more than 30 subclasses to better characterize complex variable stars. We also plan to investigate alternative model architectures and improved contrastive objectives to strengthen representation quality, ultimately developing a large-scale multitask model for heterogeneous time-domain survey data.

\section*{Acknowledgements}
We thank the anonymous referee for the very helpful comments. This work was supported by the National Natural Science Foundation of China (NSFC) through grants 12322306, 12173047, 12373028, 12233009 and 12133002. X. Chen and S. Wang acknowledge support from the Youth Innovation Promotion Association of the Chinese Academy of Sciences (CAS, No. 2022055 and 2023065). This work used data collected by TESS, ZTF and Gaia. Funding for TESS is provided by NASA's Science Mission Directorate. TESS data in this paper were obtained from the Mikulski Archive for Space Telescopes (MAST) at the Space Telescope Science Institute. ZTF is supported by the National Science Foundation under grant Nos. AST-1440341 and AST-2034437. This work has made use of data from the European Space Agency (ESA) mission \textit{Gaia} (\url{https://www.cosmos.esa.int/gaia}), processed by the Gaia Data
Processing and Analysis Consortium (DPAC, \url{https://www.cosmos.esa.int/web/gaia/dpac/consortium}). Funding for the DPAC has been provided by national institutions, in particular the institutions participating in the Gaia Multilateral Agreement.

\software{Astropy \citep{astropy2,astropy1}, Pandas \citep{pandas}, NumPy \citep{NumPy}, Matplotlib \citep{matplotlib}, PyTorch \citep{pytorch}, Transformers \citep{transformers}, Accelerate \citep{accelerate}, Datasets \citep{datasets}}

\bibliography{main}{}
\bibliographystyle{aasjournal}

\appendix
\restartappendixnumbering
\section{Supporting Figures and Tables}

Tables~\ref{tab:PerClassAll} present the detailed per-class classification performance of StarCLR on the TESS, ZTF, and Gaia test sets, including uncertainty estimates from multiple runs.

\begin{longtable}{lcccc}
\caption{Per-class classification performance of the StarCLR model on the TESS, ZTF, and Gaia test sets.}
\label{tab:PerClassAll}\\

\toprule
Type & Precision (\%) & Recall (\%) & $F_{1}$ (\%) & Number \\
\midrule
\endfirsthead

\toprule
Type & Precision (\%) & Recall (\%) & $F_{1}$ (\%) & Number \\
\midrule
\endhead

\midrule
\multicolumn{5}{r}{\emph{Continued on next page}} \\
\midrule
\endfoot

\bottomrule
\endlastfoot

\multicolumn{5}{c}{\textbf{TESS test set}} \\
\midrule
BCEP  & $72.02 \pm 2.28$ & $87.72 \pm 1.52$ & $79.07 \pm 0.78$ & 38 \\
DCEP  & $84.52 \pm 4.55$ & $90.09 \pm 3.12$ & $87.11 \pm 1.36$ & 111 \\
DCEPS & $71.28 \pm 4.80$ & $75.00 \pm 16.24$ & $72.17 \pm 7.16$ & 32 \\
DSCT  & $97.90 \pm 1.00$ & $99.12 \pm 0.21$ & $98.51 \pm 0.56$ & 2,427 \\
EA    & $98.07 \pm 0.30$ & $97.63 \pm 1.08$ & $97.85 \pm 0.45$ & 2,410 \\
EB    & $90.79 \pm 2.33$ & $85.79 \pm 8.89$ & $88.13 \pm 5.88$ & 373 \\
EW    & $95.94 \pm 3.11$ & $96.26 \pm 1.00$ & $96.09 \pm 1.87$ & 1,274 \\
GCAS  & $83.57 \pm 7.22$ & $36.67 \pm 5.77$ & $50.57 \pm 4.15$ & 30 \\
GDOR  & $87.96 \pm 3.07$ & $91.53 \pm 7.39$ & $89.67 \pm 5.12$ & 59 \\
HADS  & $88.68 \pm 6.34$ & $92.86 \pm 1.02$ & $90.63 \pm 3.10$ & 98 \\
ROT   & $92.63 \pm 1.48$ & $95.12 \pm 0.27$ & $93.85 \pm 0.64$ & 369 \\
RRab  & $97.04 \pm 0.55$ & $99.19 \pm 0.93$ & $98.10 \pm 0.19$ & 165 \\
RRcd  & $97.85 \pm 1.43$ & $89.29 \pm 2.06$ & $93.36 \pm 1.01$ & 84 \\
SPB   & $93.19 \pm 2.78$ & $87.96 \pm 2.89$ & $90.47 \pm 2.08$ & 72 \\
T2CEP & $68.06 \pm 6.36$ & $36.36 \pm 9.09$ & $46.56 \pm 6.33$ & 11 \\
UG    & $70.88 \pm 3.02$ & $72.65 \pm 2.96$ & $71.73 \pm 2.57$ & 39 \\
UV    & $95.58 \pm 1.62$ & $97.79 \pm 1.19$ & $96.66 \pm 0.56$ & 256 \\
YSO   & $77.23 \pm 2.53$ & $59.05 \pm 11.55$ & $66.61 \pm 8.64$ & 70 \\
\midrule
Macro-$F_1$ (\%) &  &  & $84.35 \pm 1.32$ &  \\
Micro-$F_1$ (\%) &  &  & $96.46 \pm 0.27$ &  \\
Total number &  &  &  & 7,918 \\
\midrule\midrule

\multicolumn{5}{c}{\textbf{ZTF test set}} \\
\midrule
BYDra & $79.31 \pm 1.95$ & $74.14 \pm 2.22$ & $76.60 \pm 0.57$ & 1,373 \\
CEP   & $91.80 \pm 0.44$ & $94.17 \pm 0.97$ & $92.97 \pm 0.32$ & 103 \\
CEPII & $97.22 \pm 4.81$ & $46.67 \pm 2.31$ & $63.06 \pm 3.12$ & 25 \\
DSCT  & $95.24 \pm 0.34$ & $96.30 \pm 0.26$ & $95.77 \pm 0.05$ & 1,253 \\
EA    & $83.37 \pm 1.55$ & $84.70 \pm 1.33$ & $84.01 \pm 0.14$ & 4,439 \\
EW    & $95.43 \pm 0.06$ & $95.66 \pm 0.18$ & $95.54 \pm 0.07$ & 31,629 \\
Mira  & $97.02 \pm 0.62$ & $93.87 \pm 0.93$ & $95.42 \pm 0.20$ & 647 \\
RR    & $98.43 \pm 0.27$ & $96.58 \pm 0.15$ & $97.49 \pm 0.07$ & 2,337 \\
RRc   & $89.29 \pm 0.75$ & $87.76 \pm 0.69$ & $88.52 \pm 0.25$ & 937 \\
RSCVN & $83.52 \pm 0.66$ & $81.32 \pm 0.46$ & $82.40 \pm 0.14$ & 3,815 \\
SR    & $91.50 \pm 0.27$ & $97.20 \pm 0.39$ & $94.26 \pm 0.07$ & 1,878 \\
\midrule
Macro-$F_1$ (\%) &  &  & $87.82 \pm 0.29$ &  \\
Micro-$F_1$ (\%) &  &  & $92.83 \pm 0.06$ &  \\
Total number &  &  &  & 48,436 \\
\midrule\midrule

\multicolumn{5}{c}{\textbf{Gaia test set}} \\
\midrule
ACV\textbar CP\textbar MCP\textbar ROAM\textbar ROAP\textbar SXARI & $97.48 \pm 0.35$ & $98.45 \pm 0.77$ & $97.96 \pm 0.20$ & 601 \\
ACYG        & $100.00 \pm 0.00$ & $92.86 \pm 0.00$ & $96.30 \pm 0.00$ & 14 \\
AGN         & $99.10 \pm 0.33$ & $99.41 \pm 0.30$ & $99.25 \pm 0.05$ & 14,176 \\
BCEP        & $86.20 \pm 1.84$ & $92.91 \pm 1.23$ & $89.42 \pm 1.52$ & 47 \\
BE\textbar GCAS\textbar SDOR\textbar WR & $98.82 \pm 0.14$ & $98.24 \pm 0.00$ & $98.53 \pm 0.07$ & 797 \\
CEP         & $99.47 \pm 0.30$ & $98.98 \pm 0.52$ & $99.22 \pm 0.11$ & 688 \\
CV          & $99.54 \pm 0.30$ & $99.27 \pm 0.11$ & $99.40 \pm 0.17$ & 503 \\
DSCT\textbar GDOR\textbar SXPHE & $97.18 \pm 1.23$ & $96.36 \pm 1.06$ & $96.76 \pm 0.17$ & 2,004 \\
ECL         & $99.49 \pm 0.16$ & $99.43 \pm 0.18$ & $99.46 \pm 0.02$ & 16,178 \\
ELL         & $99.97 \pm 0.05$ & $99.96 \pm 0.04$ & $99.96 \pm 0.01$ & 2,272 \\
EP          & $100.00 \pm 0.00$ & $96.12 \pm 1.34$ & $98.02 \pm 0.70$ & 43 \\
LPV         & $99.86 \pm 0.00$ & $99.99 \pm 0.01$ & $99.92 \pm 0.00$ & 21,368 \\
MICROLENSING & $100.00 \pm 0.00$ & $76.19 \pm 8.25$ & $86.32 \pm 5.18$ & 7 \\
RCB         & $100.00 \pm 0.00$ & $86.67 \pm 23.09$ & $91.67 \pm 14.43$ & 5 \\
RR          & $99.48 \pm 0.13$ & $99.81 \pm 0.07$ & $99.64 \pm 0.06$ & 2,294 \\
RS          & $99.79 \pm 0.05$ & $99.53 \pm 0.28$ & $99.66 \pm 0.16$ & 1,900 \\
S           & $99.94 \pm 0.02$ & $99.80 \pm 0.01$ & $99.87 \pm 0.02$ & 8,110 \\
SDB         & $96.84 \pm 0.03$ & $98.92 \pm 1.08$ & $97.87 \pm 0.54$ & 93 \\
SN          & $98.65 \pm 0.01$ & $99.09 \pm 0.39$ & $98.87 \pm 0.20$ & 147 \\
SOLAR\_LIKE & $99.72 \pm 0.00$ & $99.56 \pm 0.11$ & $99.64 \pm 0.05$ & 1,070 \\
SPB         & $78.46 \pm 7.04$ & $68.75 \pm 2.08$ & $73.13 \pm 2.16$ & 48 \\
SYST        & $46.67 \pm 5.77$ & $5.38 \pm 1.86$ & $9.53 \pm 3.01$ & 31 \\
WD          & $96.18 \pm 1.85$ & $94.34 \pm 0.00$ & $95.24 \pm 0.91$ & 53 \\
YSO         & $99.81 \pm 0.02$ & $99.80 \pm 0.00$ & $99.81 \pm 0.01$ & 3,021 \\
\midrule
Macro-$F_1$ (\%) &  &  & $92.73 \pm 0.46$ &  \\
Micro-$F_1$ (\%) &  &  & $99.49 \pm 0.01$ &  \\
Total number &  &  &  & 75,470 \\
\end{longtable}



\end{document}